\documentclass[prd,aps,floatfix,superscriptaddress,twocolumn]{revtex4-2}
\usepackage{amsfonts,amsmath,amssymb,amsthm}
\usepackage{bm}
\usepackage{booktabs}
\usepackage{braket} 
\usepackage{cases}
\usepackage{color}
\usepackage{dcolumn}
\usepackage{epsfig}
\usepackage{epstopdf}   
\usepackage{graphicx}
\usepackage[colorlinks,citecolor=blue,anchorcolor=red,menucolor=red,linkcolor=red,filecolor=red,runcolor=red,urlcolor=blue,frenchlinks=red]{hyperref}
\usepackage{indentfirst}
\usepackage{latexsym}
\usepackage{longtable}
\usepackage{multirow}
\usepackage{rotating}
\usepackage{slashed}  
\usepackage{subfigure}
\usepackage{youngtab}

\allowdisplaybreaks[4]
\begin{document}
\title{Properties of $Q^{5}q$ dibaryons}
\author{Xin-Zhen Weng}
\email{xinzhenweng@mail.tau.ac.il}
\affiliation{School of Physics and Astronomy, Tel Aviv University, Tel Aviv 6997801, Israel}
\date{\today}
\begin{abstract}

We investigate heavy flavor dibaryons with five heavy quarks $Q$ ($Q=\{c,b\}$) and one light quark $q$ ($q=\{u,d,s\}$), namely the $Q^{5}q$ dibaryons.
In the framework of an extended chromomagnetic model, we systematically study the mass spectrum of
these dibaryons.
We find no stable state below the corresponding baryon-baryon thresholds.
In addition to the analysis of the masses, we also study their two body decay properties by estimating the relative width ratios of the decay channels.
We hope our study will be of help for future experiments.
%

\end{abstract}

\maketitle
\thispagestyle{empty} 

\section{Introduction}
\label{Sec:Introduction}

In the last few decades, lots of exotic states which cannot be explained as the conventional mesons or baryons were observed in experiment.
In 2003, the Belle Collaboration observed the $X(3872)$ state in the exclusive $B^{\pm}{\to}K^{\pm}\pi^{+}\pi^{-}J/\psi$ decays~\cite{Belle:2003nnu}.
It was later confirmed by the CDF~\cite{CDF:2003cab}, D0~\cite{D0:2004zmu}, BABAR~\cite{BaBar:2004oro}, LHCb~\cite{LHCb:2011zzp}, CMS~\cite{CMS:2013fpt}, and BESIII~\cite{BESIII:2013fnz} Collaborations.
Its quantum numbers are $I^{G}J^{PC}=0^{+}1^{++}$~\cite{Workman:2022ynf}.
Since then, lots of charmoniumlike and bottomoniumlike $XYZ$ states have been observed in experiments~\cite{Workman:2022ynf}.
In 2020, A narrow structure, $X(6900)$, matching the lineshape of a resonance and a broad structure next to the di-$J/\psi$ mass threshold were found by the LHCb collaboration in the $J/\psi$-pair invariant mass spectrum~\cite{LHCb:2020bwg}.
These structures were later confirmed by the ATLAS collaboration~\cite{ATLAS:2023bft}.
Recently, the CMS collaboration confirmed the $X(6900)$ and observed two additional states with mass $6552~\text{MeV}$ and $7287~\text{MeV}$~\cite{CMS:2023owd}.
These states could be the $cc\bar{c}\bar{c}$ tetraquarks.
%
Besides the hidden flavor states, many open heavy flavor states were observed as well.
In 2020, the LHCb Collaboration observed two resonances in the $D^{-}K^{+}$ channel~\cite{LHCb:2020bls,LHCb:2020pxc}.
They are the $X_{0/1}(2900)$ states which consist of only one heavy charm quark.
In the next year, a narrow doubly-charmed tetraquark $T_{cc}^{+}$ was observed in the $D^{0}D^{0}\pi^{+}$ mass spectrum, just below the $D^{*+}D^{0}$ threshold~\cite{LHCb:2021vvq,LHCb:2021auc}.
There are also great experimental progresses in the pentaquark sector.
In 2015, the LHCb Collaboration observed the $P_{c}(4380)$ and $P_{c}(4450)$ states in the ${J/\psi}p$ mass spectrum~\cite{LHCb:2015yax}.
An later study found that the $P_{c}(4450)^{+}$ is actually composed of two
narrow resonances, $P_{c}(4440)^{+}$ and $P_{c}(4457)^{+}$.
And they also found a lower mass state $P_{c}(4312)^{+}$~\cite{LHCb:2019kea}.
In 2021, the LHCb collaboration found an additional state $P_{c}(4337)$ in the same channel~\cite{LHCb:2021chn}.
These states have a minimal quark content of $uudc\bar{c}$.
Moreover, the LHCb collaboration also observed their strange partners, $P_{cs}(4338)$~\cite{LHCb:2022ogu} and $P_{cs}(4459)$~\cite{LHCb:2020jpq}, in the $J/\psi\Lambda$ channel.
More details of the experimental as well as theoretical progresses can be found in review Refs.~\cite{Chen:2016qju,Chen:2016spr,Esposito:2016noz,Lebed:2016hpi,Ali:2017jda,Karliner:2017qhf,Olsen:2017bmm,Guo:2017jvc,Yuan:2018inv,Liu:2019zoy,Brambilla:2019esw,Bicudo:2022cqi,Chen:2022asf,Meng:2022ozq} and references therein.

The study of the six-quark systems has also been pursued for very long time.
In 1932, Urey, Brickwedde and Murphy discovered the deuteron in the atomic spectrum of hydrogen gas in a discharge tube~\cite{Urey:1932gik,Urey:1932pvy}.
It is a very loosely bound $p^{+}n^{0}$ molecule with a binding energy of only $2.2~\text{MeV}$.
Its quantum numbers are $IJ^{P}=01^{+}$.
%
Another dibaryon candidate is the $d^{*}(2380)$.
In Ref.~\cite{Bashkanov:2008ih}, the CELSIUS/WASA Collaboration studied the double-pionic fusion reaction $pn{\rightarrow}d\pi^{0}\pi^{0}$.
They observed a structure with with the mass and width of the $s$-channel resonance being $M\approx2.36~\text{GeV}$ and $\Gamma\approx80~\text{MeV}$.
This resonance was later confirmed by experiments performed by the WASA-at-COSY Collaboration~\cite{Bashkanov:2008ih,WASA-at-COSY:2011bjg,WASA-at-COSY:2012seb,WASA-at-COSY:2014dmv,WASA-at-COSY:2014lmt,WASA-at-COSY:2014qkg}.
Its quantum numbers are $IJ^{P}=03^{+}$.
This state was extensively explored theoretically~\cite{Dyson:1964xwa,Kamae:1976at,Chen:2014vha,Dong:2017mio,Lu:2017uey,Bashkanov:2019mbz,Gal:2019xju,Shi:2019dpo,Gongyo:2020pyy,Kim:2020rwn,Kukulin:2020gee,Platonova:2021trw,Pan:2023wrm}.
For example, such a system was studied by Dyson and Xuong in as early as 1964, with an impressive prediction of its mass around $2350~\text{MeV}$~\cite{Dyson:1964xwa}.
In Ref.~\cite{Kamae:1976at}, Kamae and Fujita studied the $\Delta\Delta$ system by a non-relativistic one-boson-exchange (OBE) potential model and found that the two $\Delta$ isobars are bound by about $100~\text{MeV}$.
Besides the light systems, dibaryons with heavy flavor(s) have also been extensively studied in various models, such as 
the potential quark model~\cite{Leandri:1993zg,Leandri:1995zm,Wang:1995bg,Pepin:1998ih,Gal:2014jza,Carames:2015sya,Vijande:2016nzk,Park:2016mez,Richard:2020zxb,Garcilazo:2020acl,Huang:2020bmb,Xia:2021tof,Alcaraz-Pelegrina:2022fsi,Weng:2022ohh,Lu:2022myk,Liu:2022rzu,Deng:2023zlx},
the OBE potential model~\cite{Meguro:2011nr,Chen:2018pzd,Maeda:2018xcl,Yang:2019rgw,Pan:2020xek,Ling:2021asz,Kong:2022rvd,Song:2023cst},
the QCD sum rules~\cite{Wang:2019gal,Wang:2021qmn},
the effective field theory~\cite{Lu:2017dvm,Song:2020isu,Chen:2021cfl},
the lattice QCD~\cite{Junnarkar:2019equ,Lyu:2021qsh,Junnarkar:2022yak,Mathur:2022ovu},
and the dispersion relation technique~\cite{Gerasyuta:2011yg,Gerasyuta:2011zx}.
More details can be found in the recent review Ref.~\cite{Clement:2016vnl} and references therein.

For most of these systems, the exchange of a light mesons may be important.
If so, it is difficult to determine whether they are loosely bound molecular states or compact dibaryons.
However, for the $Q^{5}q$ and $Q^{6}$ systems, a molecular configuration requires exchange of a heavy meson, which is highly suppressed in the typical scale of the molecule ($\sim\mathrm{fm}$).
Thus the molecular interpretation is not favored.
If these dibaryons are observed in experiment, they are very likely compact states.
%

\begin{table*}
\centering
\caption{Parameters of the $qq$ pairs for baryons~\cite{Weng:2018mmf} (in units of $\text{MeV}$).}
\label{table:parameter:qq}
\begin{tabular}{lcccccccccccc}
\toprule[1pt]
\toprule[1pt]
Parameter&$m_{nn}^{b}$&$m_{ns}^{b}$&$m_{ss}^{b}$&$m_{nc}^{b}$&$m_{sc}^{b}$&$m_{cc}^{b}$&$m_{nb}^{b}$&$m_{sb}^{b}$&$m_{cb}^{b}$&$m_{b{b}}^{b}$\\
Value&$724.85$&$906.65$&$1049.36$&$2079.96$&$2183.68$&$3171.51$&$5412.25$&$5494.80$&$6416.07$&$9529.57$\\
Parameter&$v_{n{n}}^{b}$&$v_{n{s}}^{b}$&$v_{ss}^{b}$&$v_{n{c}}^{b}$&$v_{s{c}}^{b}$&$v_{c{c}}^{b}$&$v_{n{b}}^{b}$&$v_{s{b}}^{b}$&$v_{c{b}}^{b}$&$v_{b{b}}^{b}$\\
Value&$305.34$&$212.75$&$195.30$&$62.81$&$70.63$&$56.75$&$19.92$&$8.47$&$31.45$&$30.65$\\
\bottomrule[1pt]
\bottomrule[1pt]
\end{tabular}
\end{table*}

For these compact systems, the interaction is provided by gluon exchange and string confinement.
The interactions include spin-independent Coulomb-type interaction and linear confinement, plus higher order terms such as spin-spin chromomagnetic interaction, tensor interaction, and spin-orbit interaction.
In this work, we only consider the $S$-wave states, thus the tensor and spin-orbit interactions can be neglected.
Then we can use the extended chromomagnetic model~\cite{Hogaasen:2013nca,Weng:2018mmf,Liu:2019zoy}.
In this model, the hadron masses receive contributions from effective quark masses, color interaction and chromomagnetic interaction.
This simplified model provides good account of all $S$-wave mesons and baryons~\cite{Weng:2018mmf}.
In Ref.~\cite{Weng:2022ohh}, we have used this model to investigate the fully heavy dibaryons.
Here, we estimate the mass spectrum and decay property of the $Q^{5}q$ dibaryons within the same model.

The paper is organized as follows.
In Sec.~\ref{Sec:Model}, we introduce the extended chromomagnetic model and present the wave function bases of the $Q^{5}q$ dibaryons.
In Sec.~\ref{Sec:Result}, we calculate and discuss the numerical results.
We conclude in Sec.~\ref{Sec:Conclusion}.
%

\section{The Extended Chromomagnetic Model}
\label{Sec:Model}

In the quark model, the Hamiltonian of the $S$-wave hadron reads~\cite{Chan:1978nk,Hogaasen:2013nca}
\begin{equation}\label{eqn:ECM}
H
=
\sum_{i}m_{i}+H_{\text{CE}}+H_{\text{CM}}\,,
\end{equation}
where $m_i$ is the effective mass of $i$th quark.
$H_{\text{CE}}$ is the colorelectric (CE) interaction
\begin{equation}\label{eqn:ECM:CE}
H_{\text{CE}}
=
-
\sum_{i<j}
a_{ij}
\bm{F}_{i}\cdot\bm{F}_{j}\,,
\end{equation}
and $H_{\text{CM}}$ is the chromomagnetic (CM) interaction
\begin{equation}\label{eqn:ECM:CM}
H_{\text{CM}}
=
-
\sum_{i<j}
v_{ij}
\bm{S}_{i}\cdot\bm{S}_{j}
\bm{F}_{i}\cdot\bm{F}_{j}\,,
\end{equation}
Here, $a_{ij}$ and $v_{ij}\propto\braket{\alpha(\mathbf{r})\delta^{3}(\mathbf{r})}/m_{i}m_{j}$ are effective coupling constants which depend on the constituent quark masses and the spatial wave function.
$\bm{S}_{i}=\bm{\sigma}_i/2$ and $\bm{F}_{i}={\bm{\lambda}}_i/2$ are the quark spin and  color operators.

Since
\begin{align}\label{eqn:m+color=color}
&
\sum_{i<j}
\left(m_i+m_j\right)
\bm{F}_{i}\cdot\bm{F}_{j}
\notag\\
={}&
\left(\sum_{i}m_{i}\bm{F}_i\right)
\cdot
\left(\sum_{i}\bm{F}_{i}\right)
-
\frac{4}{3}
\sum_{i}
m_{i}\,,
\end{align}
and the total color operator $\sum_i\bm{F}_i$ nullifies any color-singlet physical state, we can rewrite the Hamiltonian as~\cite{Weng:2018mmf}
\begin{equation}\label{eqn:hamiltonian:final}
	H=
	-\frac{3}{4}
	\sum_{i<j}m_{ij}V^{\text{C}}_{ij}
	-
	\sum_{i<j}v_{ij}V^{\text{CM}}_{ij} \,,
\end{equation}
by introducing the quark pair mass parameter
\begin{equation}\label{eqn:para:color+m}
	m_{ij}
	=
	\left(m_i+m_j\right)
	+
	\frac{4}{3}
	a_{ij}\,,
\end{equation}
where $V^{\text{C}}_{ij}\equiv\bm{F}_{i}\cdot\bm{F}_{j}$ and $V^{\text{CM}}_{ij}\equiv\bm{S}_{i}\cdot\bm{S}_{j}\bm{F}_{i}\cdot\bm{F}_{j}$ are the color and CM interactions between quarks.
The model parameters $m_{ij}$ and $v_{ij}$ can be fitted from conventional mesons and baryons.
Their values are listed in Table~\ref{table:parameter:qq}.
In this work, we use these parameters to estimate the masses of the $S$-wave $Q^{5}q$ dibaryons.

To obtain the mass spectra of the dibaryons, we need to construct the wave functions.
In principle, the total wave function is a direct product of the spatial, flavor, color and spin wave functions.
In Table~\ref{table:wavefunc:color+spin:12x3x4x5x6:12x3x45x6}, we construct the color and spin wave functions of the dibaryons in the $\left\{\left[\left(q_{1}q_{2}{\otimes}q_{3}\right){\otimes}q_{4}\right]{\otimes}q_{5}\right\}{\otimes}q_{6}$ and $\left[\left(q_{1}q_{2}{\otimes}q_{3}\right){\otimes}q_{4}q_{5}\right]{\otimes}q_{6}$ configurations.
In this work, we consider the ground state and assume that the spatial wave function is totally symmetric.
Then we need to construct the totally anti-symmetric $\text{flavor}\otimes\text{color}\otimes\text{spin}$ wave functions.
The total wave functions of the $Q^{5}q$ dibaryons are listed in Tables~\ref{table:wavefunc:total}.
More details can be found in Ref.~\cite{Weng:2022ohh}.
Diagonalizing the Hamiltonian in these bases, we can obtain the masses and eigenvectors of the $Q^{5}q$ dibaryons.
%

\begin{table*}[htbp]
\centering
\caption{The dibaryon color and spin wave functions in the $\left\{\left[\left(q_{1}q_{2}{\otimes}q_{3}\right){\otimes}q_{4}\right]{\otimes}q_{5}\right\}{\otimes}q_{6}$ and $\left[\left(q_{1}q_{2}{\otimes}q_{3}\right){\otimes}q_{4}q_{5}\right]{\otimes}q_{6}$ configurations. The superscripts $\{1,3,\bar{3},6,8\}$ are color representations, and the subscripts $\{0,1/2,1,3/2,2,5/2,3\}$ are spins.}
\label{table:wavefunc:color+spin:12x3x4x5x6:12x3x45x6}
\begin{tabular}{cccccccccccccccccccccccc}
\toprule[1pt]
\toprule[1pt]
&
&$\left\{\left[\left(q_{1}q_{2}{\otimes}q_{3}\right){\otimes}q_{4}\right]{\otimes}q_{5}\right\}{\otimes}q_{6}$
&$\left[\left(q_{1}q_{2}{\otimes}q_{3}\right){\otimes}q_{4}q_{5}\right]{\otimes}q_{6}$\\
\midrule[1pt]
{Color}
&&$\phi_{\alpha1}=\ket{(\{[(q_{1}q_{2})^{6}q_{3}]^{8}q_{4}\}^{\bar{6}}q_{5})^{\bar{3}}q_{6}}^{1}$
&$\phi_{\beta1}=\ket{\{[(q_{1}q_{2})^{6}q_{3}]^{8}(q_{4}q_{5})^{6}\}^{\bar{3}}q_{6}}^{1}$\\
&&$\phi_{\alpha2}=\ket{(\{[(q_{1}q_{2})^{\bar{3}}q_{3}]^{8}q_{4}\}^{\bar{6}}q_{5})^{\bar{3}}q_{6}}^{1}$
&$\phi_{\beta2}=\ket{\{[(q_{1}q_{2})^{\bar{3}}q_{3}]^{8}(q_{4}q_{5})^{6}\}^{\bar{3}}q_{6}}^{1}$\\
&&$\phi_{\alpha3}=\ket{(\{[(q_{1}q_{2})^{6}q_{3}]^{8}q_{4}\}^{3}q_{5})^{\bar{3}}q_{6}}^{1}$
&$\phi_{\beta3}=\ket{\{[(q_{1}q_{2})^{6}q_{3}]^{8}(q_{4}q_{5})^{\bar{3}}\}^{\bar{3}}q_{6}}^{1}$\\
&&$\phi_{\alpha4}=\ket{(\{[(q_{1}q_{2})^{\bar{3}}q_{3}]^{8}q_{4}\}^{3}q_{5})^{\bar{3}}q_{6}}^{1}$
&$\phi_{\beta4}=\ket{\{[(q_{1}q_{2})^{\bar{3}}q_{3}]^{8}(q_{4}q_{5})^{\bar{3}}\}^{\bar{3}}q_{6}}^{1}$\\
&&$\phi_{\alpha5}=\ket{(\{[(q_{1}q_{2})^{\bar{3}}q_{3}]^{1}q_{4}\}^{3}q_{5})^{\bar{3}}q_{6}}^{1}$
&$\phi_{\beta5}=\ket{\{[(q_{1}q_{2})^{\bar{3}}q_{3}]^{1}(q_{4}q_{5})^{\bar{3}}\}^{\bar{3}}q_{6}}^{1}$\\
\midrule[1pt]
Spin
&$J=0$
&$\chi^{0}_{\alpha1}=\ket{(\{[(q_{1}q_{2})_{1}q_{3}]_{3/2}q_{4}\}_{1}q_{5})_{1/2}q_{6}}_{0}$
&$\chi^{0}_{\beta1}=\ket{\{[(q_{1}q_{2})_{1}q_{3}]_{3/2}(q_{4}q_{5})_{1}\}_{1/2}q_{6}}_{0}$\\
&&$\chi^{0}_{\alpha2}=\ket{(\{[(q_{1}q_{2})_{1}q_{3}]_{1/2}q_{4}\}_{1}q_{5})_{1/2}q_{6}}_{0}$
&$\chi^{0}_{\beta2}=\ket{\{[(q_{1}q_{2})_{1}q_{3}]_{1/2}(q_{4}q_{5})_{1}\}_{1/2}q_{6}}_{0}$\\
&&$\chi^{0}_{\alpha3}=\ket{(\{[(q_{1}q_{2})_{0}q_{3}]_{1/2}q_{4}\}_{1}q_{5})_{1/2}q_{6}}_{0}$
&$\chi^{0}_{\beta3}=\ket{\{[(q_{1}q_{2})_{0}q_{3}]_{1/2}(q_{4}q_{5})_{1}\}_{1/2}q_{6}}_{0}$\\
&&$\chi^{0}_{\alpha4}=\ket{(\{[(q_{1}q_{2})_{1}q_{3}]_{1/2}q_{4}\}_{0}q_{5})_{1/2}q_{6}}_{0}$
&$\chi^{0}_{\beta4}=\ket{\{[(q_{1}q_{2})_{1}q_{3}]_{1/2}(q_{4}q_{5})_{0}\}_{1/2}q_{6}}_{0}$\\
&&$\chi^{0}_{\alpha5}=\ket{(\{[(q_{1}q_{2})_{0}q_{3}]_{1/2}q_{4}\}_{0}q_{5})_{1/2}q_{6}}_{0}$
&$\chi^{0}_{\beta5}=\ket{\{[(q_{1}q_{2})_{0}q_{3}]_{1/2}(q_{4}q_{5})_{0}\}_{1/2}q_{6}}_{0}$\\
&$J=1$
&$\chi^{1}_{\alpha1}=\ket{(\{[(q_{1}q_{2})_{1}q_{3}]_{3/2}q_{4}\}_{2}q_{5})_{3/2}q_{6}}_{1}$
&$\chi^{1}_{\beta1}=\ket{\{[(q_{1}q_{2})_{1}q_{3}]_{3/2}(q_{4}q_{5})_{1}\}_{3/2}q_{6}}_{1}$\\
&&$\chi^{1}_{\alpha2}=\ket{(\{[(q_{1}q_{2})_{1}q_{3}]_{3/2}q_{4}\}_{1}q_{5})_{3/2}q_{6}}_{1}$
&$\chi^{1}_{\beta2}=\ket{\{[(q_{1}q_{2})_{1}q_{3}]_{3/2}(q_{4}q_{5})_{0}\}_{3/2}q_{6}}_{1}$\\
&&$\chi^{1}_{\alpha3}=\ket{(\{[(q_{1}q_{2})_{1}q_{3}]_{1/2}q_{4}\}_{1}q_{5})_{3/2}q_{6}}_{1}$
&$\chi^{1}_{\beta3}=\ket{\{[(q_{1}q_{2})_{1}q_{3}]_{1/2}(q_{4}q_{5})_{1}\}_{3/2}q_{6}}_{1}$\\
&&$\chi^{1}_{\alpha4}=\ket{(\{[(q_{1}q_{2})_{0}q_{3}]_{1/2}q_{4}\}_{1}q_{5})_{3/2}q_{6}}_{1}$
&$\chi^{1}_{\beta4}=\ket{\{[(q_{1}q_{2})_{0}q_{3}]_{1/2}(q_{4}q_{5})_{1}\}_{3/2}q_{6}}_{1}$\\
&&$\chi^{1}_{\alpha5}=\ket{(\{[(q_{1}q_{2})_{1}q_{3}]_{3/2}q_{4}\}_{1}q_{5})_{1/2}q_{6}}_{1}$
&$\chi^{1}_{\beta5}=\ket{\{[(q_{1}q_{2})_{1}q_{3}]_{3/2}(q_{4}q_{5})_{1}\}_{1/2}q_{6}}_{1}$\\
&&$\chi^{1}_{\alpha6}=\ket{(\{[(q_{1}q_{2})_{1}q_{3}]_{1/2}q_{4}\}_{1}q_{5})_{1/2}q_{6}}_{1}$
&$\chi^{1}_{\beta6}=\ket{\{[(q_{1}q_{2})_{1}q_{3}]_{1/2}(q_{4}q_{5})_{1}\}_{1/2}q_{6}}_{1}$\\
&&$\chi^{1}_{\alpha7}=\ket{(\{[(q_{1}q_{2})_{0}q_{3}]_{1/2}q_{4}\}_{1}q_{5})_{1/2}q_{6}}_{1}$
&$\chi^{1}_{\beta7}=\ket{\{[(q_{1}q_{2})_{0}q_{3}]_{1/2}(q_{4}q_{5})_{1}\}_{1/2}q_{6}}_{1}$\\
&&$\chi^{1}_{\alpha8}=\ket{(\{[(q_{1}q_{2})_{1}q_{3}]_{1/2}q_{4}\}_{0}q_{5})_{1/2}q_{6}}_{1}$
&$\chi^{1}_{\beta8}=\ket{\{[(q_{1}q_{2})_{1}q_{3}]_{1/2}(q_{4}q_{5})_{0}\}_{1/2}q_{6}}_{1}$\\
&&$\chi^{1}_{\alpha9}=\ket{(\{[(q_{1}q_{2})_{0}q_{3}]_{1/2}q_{4}\}_{0}q_{5})_{1/2}q_{6}}_{1}$
&$\chi^{1}_{\beta9}=\ket{\{[(q_{1}q_{2})_{0}q_{3}]_{1/2}(q_{4}q_{5})_{0}\}_{1/2}q_{6}}_{1}$\\
&$J=2$
&$\chi^{2}_{\alpha1}=\ket{(\{[(q_{1}q_{2})_{1}q_{3}]_{3/2}q_{4}\}_{2}q_{5})_{5/2}q_{6}}_{2}$
&$\chi^{2}_{\beta1}=\ket{\{[(q_{1}q_{2})_{1}q_{3}]_{3/2}(q_{4}q_{5})_{1}\}_{5/2}q_{6}}_{2}$\\
&&$\chi^{2}_{\alpha2}=\ket{(\{[(q_{1}q_{2})_{1}q_{3}]_{3/2}q_{4}\}_{2}q_{5})_{3/2}q_{6}}_{2}$
&$\chi^{2}_{\beta2}=\ket{\{[(q_{1}q_{2})_{1}q_{3}]_{3/2}(q_{4}q_{5})_{1}\}_{3/2}q_{6}}_{2}$\\
&&$\chi^{2}_{\alpha3}=\ket{(\{[(q_{1}q_{2})_{1}q_{3}]_{3/2}q_{4}\}_{1}q_{5})_{3/2}q_{6}}_{2}$
&$\chi^{2}_{\beta3}=\ket{\{[(q_{1}q_{2})_{1}q_{3}]_{3/2}(q_{4}q_{5})_{0}\}_{3/2}q_{6}}_{2}$\\
&&$\chi^{2}_{\alpha4}=\ket{(\{[(q_{1}q_{2})_{1}q_{3}]_{1/2}q_{4}\}_{1}q_{5})_{3/2}q_{6}}_{2}$
&$\chi^{2}_{\beta4}=\ket{\{[(q_{1}q_{2})_{1}q_{3}]_{1/2}(q_{4}q_{5})_{1}\}_{3/2}q_{6}}_{2}$\\
&&$\chi^{2}_{\alpha5}=\ket{(\{[(q_{1}q_{2})_{0}q_{3}]_{1/2}q_{4}\}_{1}q_{5})_{3/2}q_{6}}_{2}$
&$\chi^{2}_{\beta5}=\ket{\{[(q_{1}q_{2})_{0}q_{3}]_{1/2}(q_{4}q_{5})_{1}\}_{3/2}q_{6}}_{2}$\\
&$J=3$
&$\chi^{3}_{\alpha1}=\ket{(\{[(q_{1}q_{2})_{1}q_{3}]_{3/2}q_{4}\}_{2}q_{5})_{5/2}q_{6}}_{3}$
&$\chi^{3}_{\beta1}=\ket{\{[(q_{1}q_{2})_{1}q_{3}]_{3/2}(q_{4}q_{5})_{1}\}_{5/2}q_{6}}_{3}$\\
\bottomrule[1pt]
\bottomrule[1pt]
\end{tabular}
\end{table*}
%
\begin{table*}[htbp]
\centering
\caption{The total wave functions of $Q^{5}q$ dibaryons. The color ($\phi$) and spin ($\chi$) wave functions are defined in Table~\ref{table:wavefunc:color+spin:12x3x4x5x6:12x3x45x6}.}
\label{table:wavefunc:total}
\begin{tabular}{cllccccccccccccccccccccc}
\toprule[1pt]
\toprule[1pt]
&\multicolumn{2}{c}{$\varphi_{\text{A}}=\{c^{5}q,b^{5}q\}$}\\
\midrule[1pt]
$J=0$
&\multicolumn{2}{c}{$\Psi_{\text{A}1}^{0}=\frac{1}{\sqrt{5}}\varphi_{\text{A}}\left(\phi_{\alpha1}\chi^{0}_{\alpha5}-\phi_{\alpha2}\chi^{0}_{\alpha4}-\phi_{\alpha3}\chi^{0}_{\alpha3}+\phi_{\alpha4}\chi^{0}_{\alpha2}-\phi_{\alpha5}\chi^{0}_{\alpha1}\right)$}\\
$J=1$
&\multicolumn{2}{c}{$\Psi_{\text{A}1}^{1}=\frac{1}{\sqrt{5}}\varphi_{\text{A}}\left(\phi_{\alpha1}\chi^{1}_{\alpha9}-\phi_{\alpha2}\chi^{1}_{\alpha8}-\phi_{\alpha3}\chi^{1}_{\alpha7}+\phi_{\alpha4}\chi^{1}_{\alpha6}-\phi_{\alpha5}\chi^{1}_{\alpha5}\right)$}\\
\midrule[1pt]
&\multicolumn{1}{c}{$\varphi_{\text{B}}=\{c^{4}bq,b^{4}cq\}$}
&\multicolumn{1}{c}{$\varphi_{\text{C}}=\{c^{3}b^{2}q,b^{3}c^{2}q\}$}\\
\midrule[1pt]
$J=0$
&$\Psi_{\text{B}1}^{0}=\frac{1}{\sqrt{2}}\varphi_{\text{B}}\left(\phi_{\alpha1}\chi^{0}_{\alpha5}-\phi_{\alpha2}\chi^{0}_{\alpha4}\right)$
&$\Psi_{\text{C}1}^{0}=\frac{1}{\sqrt{2}}\varphi_{\text{C}}\left(\phi_{\beta1}\chi^{0}_{\beta5}-\phi_{\beta2}\chi^{0}_{\beta4}\right)$\\
&$\Psi_{\text{B}2}^{0}=\frac{1}{\sqrt{3}}\varphi_{\text{B}}\left(\phi_{\alpha3}\chi^{0}_{\alpha3}-\phi_{\alpha4}\chi^{0}_{\alpha2}+\phi_{\alpha5}\chi^{0}_{\alpha1}\right)$
&$\Psi_{\text{C}2}^{0}=\frac{1}{\sqrt{2}}\varphi_{\text{C}}\left(\phi_{\beta3}\chi^{0}_{\beta3}-\phi_{\beta4}\chi^{0}_{\beta2}\right)$\\
&
&$\Psi_{\text{C}3}^{0}=\varphi_{\text{C}}\phi_{\beta5}\chi^{0}_{\beta1}$\\
$J=1$
&$\Psi_{\text{B}1}^{1}=\frac{1}{\sqrt{2}}\varphi_{\text{B}}\left(\phi_{\alpha1}\chi^{1}_{\alpha9}-\phi_{\alpha2}\chi^{1}_{\alpha8}\right)$
&$\Psi_{\text{C}1}^{1}=\frac{1}{\sqrt{2}}\varphi_{\text{C}}\left(\phi_{\beta1}\chi^{1}_{\beta9}-\phi_{\beta2}\chi^{1}_{\beta8}\right)$\\
&$\Psi_{\text{B}2}^{1}=\frac{1}{\sqrt{3}}\varphi_{\text{B}}\left(\phi_{\alpha3}\chi^{1}_{\alpha4}-\phi_{\alpha4}\chi^{1}_{\alpha3}+\phi_{\alpha5}\chi^{1}_{\alpha2}\right)$
&$\Psi_{\text{C}2}^{1}=\frac{1}{\sqrt{2}}\varphi_{\text{C}}\left(\phi_{\beta3}\chi^{1}_{\beta4}-\phi_{\beta4}\chi^{1}_{\beta3}\right)$\\
&$\Psi_{\text{B}3}^{1}=\frac{1}{\sqrt{3}}\varphi_{\text{B}}\left(\phi_{\alpha3}\chi^{1}_{\alpha7}-\phi_{\alpha4}\chi^{1}_{\alpha6}+\phi_{\alpha5}\chi^{1}_{\alpha5}\right)$
&$\Psi_{\text{C}3}^{1}=\frac{1}{\sqrt{2}}\varphi_{\text{C}}\left(\phi_{\beta3}\chi^{1}_{\beta7}-\phi_{\beta4}\chi^{1}_{\beta6}\right)$\\
&
&$\Psi_{\text{C}4}^{1}=\varphi_{\text{C}}\phi_{\beta5}\chi^{1}_{\beta1}$\\
&
&$\Psi_{\text{C}5}^{1}=\varphi_{\text{C}}\phi_{\beta5}\chi^{1}_{\beta5}$\\
$J=2$:
&$\Psi_{\text{B}1}^{2}=\frac{1}{\sqrt{3}}\varphi_{\text{B}}\left(\phi_{\alpha3}\chi^{2}_{\alpha5}-\phi_{\alpha4}\chi^{2}_{\alpha4}+\phi_{\alpha5}\chi^{2}_{\alpha3}\right)$
&$\Psi_{\text{C}1}^{2}=\frac{1}{\sqrt{2}}\varphi_{\text{C}}\left(\phi_{\beta3}\chi^{2}_{\beta5}-\phi_{\beta4}\chi^{2}_{\beta4}\right)$\\
&
&$\Psi_{\text{C}2}^{2}=\varphi_{\text{C}}\phi_{\beta5}\chi^{2}_{\beta1}$\\
&
&$\Psi_{\text{C}3}^{2}=\varphi_{\text{C}}\phi_{\beta5}\chi^{2}_{\beta2}$\\
$J=3$
&
&$\Psi_{\text{C}1}^{3}=\varphi_{\text{C}}\phi_{\beta5}\chi^{3}_{\beta1}$\\
\bottomrule[1pt]
\bottomrule[1pt]
\end{tabular}
\end{table*}
%

\section{Numerical results}
\label{Sec:Result}

\begin{table*}[htbp]
\centering
\caption{Masses and eigenvectors of the $cccccq$, $bbbbbq$, $ccccbq$ and $bbbbcq$ dibaryons. The masses are all in units of MeV.}
\label{table:mass:cccccq+bbbbbq+ccccbq+bbbbcq}
\begin{tabular}{ccccccccccc}
\toprule[1pt]
\toprule[1pt]
System&$J^{P}$&Mass&Eigenvector
&
System&$J^{P}$&Mass&Eigenvector\\
\midrule[1pt]
$cccccn$
&$0^{+}$
&$8599$&$\{1\}$
&
$bbbbbn$
&$0^{+}$
&$24553$&$\{1\}$\\
&$1^{+}$
&$8516$&$\{1\}$
&
&$1^{+}$
&$24526$&$\{1\}$\\
$cccccs$
&$0^{+}$
&$8711$&$\{1\}$
&
$bbbbbs$
&$0^{+}$
&$24624$&$\{1\}$\\
&$1^{+}$
&$8617$&$\{1\}$
&
&$1^{+}$
&$24612$&$\{1\}$\\
$ccccbn$
&$0^{+}$
&$11693$&$\{0.900,0.437\}$
&
$bbbbcn$
&$0^{+}$
&$21306$&$\{0.192,0.981\}$\\
&
&$11840$&$\{-0.437,0.900\}$
&
&
&$21473$&$\{0.981,-0.192\}$\\
&$1^{+}$
&$11698$&$\{0.902,-0.405,0.150\}$
&
&$1^{+}$
&$21251$&$\{0.092,-0.836,0.540\}$\\
&
&$11776$&$\{0.307,0.359,-0.882\}$
&
&
&$21303$&$\{0.045,0.546,0.837\}$\\
&
&$11806$&$\{0.303,0.841,0.448\}$
&
&
&$21448$&$\{0.995,0.052,-0.088\}$\\
&$2^{+}$
&$11765$&$\{1\}$
&
&$2^{+}$
&$21286$&$\{1\}$\\
$ccccbs$
&$0^{+}$
&$11793$&$\{0.875,0.484\}$
&
$bbbbcs$
&$0^{+}$
&$21398$&$\{0.164,0.987\}$\\
&
&$11941$&$\{-0.484,0.875\}$
&
&
&$21550$&$\{0.987,-0.164\}$\\
&$1^{+}$
&$11802$&$\{-0.872,0.467,-0.150\}$
&
&$1^{+}$
&$21341$&$\{0.072,-0.859,0.506\}$\\
&
&$11871$&$\{-0.266,-0.192,0.945\}$
&
&
&$21395$&$\{0.098,0.511,0.854\}$\\
&
&$11907$&$\{0.412,0.863,0.291\}$
&
&
&$21524$&$\{0.993,0.012,-0.121\}$\\
&$2^{+}$
&$11855$&$\{1\}$
&
&$2^{+}$
&$21383$&$\{1\}$\\
\bottomrule[1pt]
\bottomrule[1pt]
\end{tabular}
\end{table*}
%
\begin{table*}[htbp]
\centering
\caption{Masses and eigenvectors of the $cccbbq$ dibaryons. The masses are all in units of MeV.}
\label{table:mass:cccbbq}
\begin{tabular}{ccccccccccc}
\toprule[1pt]
\toprule[1pt]
System&$J^{P}$&Mass&Eigenvector&Scattering~state\\
\midrule[1pt]
$cccbbn$
&$0^{+}$
&$14887$&$\{0.087,0.925,0.369\}$\\
&
&$14981$&$\{-0.287,-0.331,0.899\}$&$\Omega_{ccc}^{*}\Xi_{bb}^{*}$~($80.8\%$)\\
&
&$15093$&$\{0.954,-0.183,0.237\}$\\
&$1^{+}$
&$14833$&$\{0.066,0.898,0.418,0.116,0.020\}$\\
&
&$14884$&$\{0.089,-0.358,0.857,-0.357,-0.038\}$\\
&
&$14953$&$\{-0.258,0.090,-0.075,-0.426,0.859\}$&$\Omega_{ccc}^{*}\Xi_{bb}$~($79.3\%$)\\
&
&$14984$&$\{0.133,0.234,-0.277,-0.823,-0.416\}$\\
&
&$15039$&$\{-0.950,0.038,0.091,-0.026,-0.294\}$\\
&$2^{+}$
&$14872$&$\{0.948,-0.314,0.048\}$\\
&
&$14962$&$\{-0.274,-0.730,0.626\}$&$\Omega_{ccc}^{*}\Xi_{bb}$~($87.0\%$)\\
&
&$14976$&$\{0.161,0.607,0.778\}$&$\Omega_{ccc}^{*}\Xi_{bb}^{*}$~($93.6\%$)\\
&$3^{+}$
&$14974$&$\{1\}$&$\Omega_{ccc}^{*}\Xi_{bb}^{*}$~($100\%$)\\
$cccbbs$
&$0^{+}$
&$14986$&$\{0.044,0.861,0.507\}$\\
&
&$15069$&$\{-0.272,-0.478,0.835\}$\\
&
&$15187$&$\{0.961,-0.175,0.214\}$\\
&$1^{+}$
&$14934$&$\{-0.051,-0.902,-0.403,-0.142,-0.033\}$\\
&
&$14984$&$\{0.123,-0.302,0.826,-0.449,-0.096\}$\\
&
&$15041$&$\{-0.263,0.038,-0.015,-0.319,0.910\}$\\
&
&$15072$&$\{0.016,0.307,-0.372,-0.822,-0.303\}$\\
&
&$15127$&$\{0.955,-0.004,-0.126,-0.024,0.266\}$\\
&$2^{+}$
&$14976$&$\{0.911,-0.411,0.033\}$\\
&
&$15051$&$\{-0.135,-0.222,0.966\}$\\
&
&$15061$&$\{-0.389,-0.884,-0.258\}$\\
&$3^{+}$
&$15053$&$\{1\}$&$\Omega_{ccc}^{*}\Omega_{bb}^{*}$~($100\%$)\\
\bottomrule[1pt]
\bottomrule[1pt]
\end{tabular}
\end{table*}
%
\begin{table*}[htbp]
\centering
\caption{Masses and eigenvectors of the $bbbccq$ dibaryons. The masses are all in units of MeV.}
\label{table:mass:bbbccq}
\begin{tabular}{ccccccccccc}
\toprule[1pt]
\toprule[1pt]
System&$J^{P}$&Mass&Eigenvector&Scattering~state\\
\midrule[1pt]
$bbbccn$
&$0^{+}$
&$18003$&$\{-0.109,0.047,0.993\}$&$\Omega_{bbb}^{*}\Xi_{cc}^{*}$~($98.6\%$)\\
&
&$18167$&$\{-0.894,-0.441,-0.077\}$\\
&
&$18280$&$\{0.434,-0.896,0.090\}$\\
&$1^{+}$
&$17941$&$\{0.068,0.043,-0.022,0.737,-0.671\}$&$\Omega_{bbb}^{*}\Xi_{cc}$~($99.3\%$)\\
&
&$18004$&$\{0.092,-0.048,0.009,-0.672,-0.733\}$&$\Omega_{bbb}^{*}\Xi_{cc}^{*}$~($98.9\%$)\\
&
&$18161$&$\{0.916,0.346,0.178,-0.007,0.101\}$\\
&
&$18233$&$\{0.385,-0.849,-0.355,0.051,0.053\}$\\
&
&$18250$&$\{0.028,0.394,-0.917,-0.046,0.008\}$\\
&$2^{+}$
&$17942$&$\{0.041,-0.893,0.449\}$&$\Omega_{bbb}^{*}\Xi_{cc}$~($99.8\%$)\\
&
&$18006$&$\{0.017,0.450,0.893\}$&$\Omega_{bbb}^{*}\Xi_{cc}^{*}$~($99.97\%$)\\
&
&$18230$&$\{0.999,0.029,-0.034\}$\\
&$3^{+}$
&$18006$&$\{1\}$&$\Omega_{bbb}^{*}\Xi_{cc}^{*}$~($100\%$)\\
$bbbccs$
&$0^{+}$
&$18109$&$\{-0.126,0.024,0.992\}$&$\Omega_{bbb}^{*}\Omega_{cc}^{*}$~($98.3\%$)\\
&
&$18251$&$\{-0.867,-0.489,-0.098\}$\\
&
&$18362$&$\{0.482,-0.872,0.083\}$\\
&$1^{+}$
&$18040$&$\{0.069,0.041,-0.018,0.735,-0.672\}$&$\Omega_{bbb}^{*}\Omega_{cc}$~($99.3\%$)\\
&
&$18110$&$\{0.096,-0.038,-0.008,-0.675,-0.730\}$&$\Omega_{bbb}^{*}\Omega_{cc}^{*}$~($98.9\%$)\\
&
&$18256$&$\{0.880,0.426,0.184,-0.010,0.101\}$\\
&
&$18323$&$\{0.448,-0.880,-0.137,0.055,0.056\}$\\
&
&$18328$&$\{0.101,0.204,-0.973,-0.018,0.030\}$\\
&$2^{+}$
&$18041$&$\{0.030,-0.893,0.450\}$&$\Omega_{bbb}^{*}\Omega_{cc}$~($99.9\%$)\\
&
&$18112$&$\{0.039,0.451,0.892\}$&$\Omega_{bbb}^{*}\Omega_{cc}^{*}$~($99.8\%$)\\
&
&$18308$&$\{0.999,0.010,-0.049\}$\\
&$3^{+}$
&$18112$&$\{1\}$&$\Omega_{bbb}^{*}\Omega_{cc}^{*}$~($100\%$)\\
\bottomrule[1pt]
\bottomrule[1pt]
\end{tabular}
\end{table*}
%

\subsection{The $cccccq$ and $bbbbbq$ systems}
\label{Sec:cccccq+bbbbbq}

The masses and eigenvectors of the $Q^{5}q$ dibaryons are listed in Tables~\ref{table:mass:cccccq+bbbbbq+ccccbq+bbbbcq}--\ref{table:mass:bbbccq}.
In this work, we assume that the $\mathrm{SU}(2)$ flavor symmetry and denote $n=\{u,d\}$.
In the following, we will use $D(QQQQQq,m,J^{P})$ to denote the $QQQQQq$ dibaryon state.
The isospin is not explicitly shown since it can be easily identified with either $q=n$ or $q=s$.

\begin{figure*}
\begin{tabular}{ccc}
\includegraphics[width=240pt]{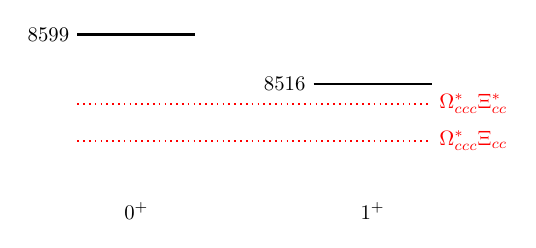}
&$\qquad$&
\includegraphics[width=240pt]{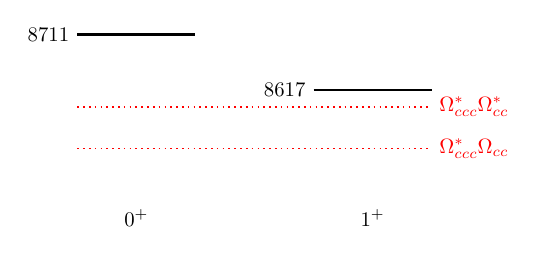}\\
(a) $cccccn$ states
&$\qquad$&
(b) $cccccs$ states\\
\includegraphics[width=240pt]{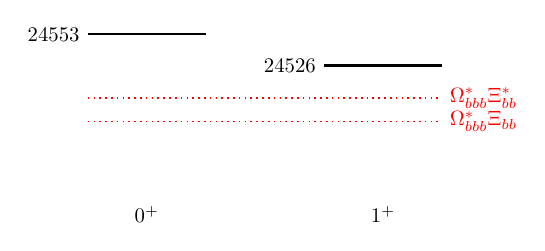}
&$\qquad$&
\includegraphics[width=240pt]{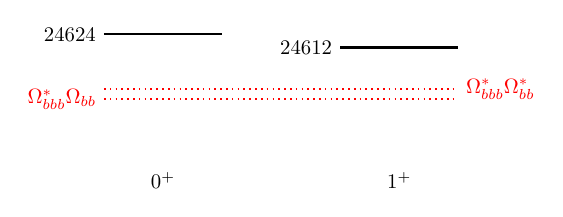}\\
(c) $bbbbbn$ states
&$\qquad$&
(d) $bbbbbs$ states\\
\end{tabular}
\caption{Masses of the $cccccq$ and $bbbbbq$ dibaryons. The dotted lines indicate various baryon-baryon thresholds, where the baryon masses are calculated in the same model~\cite{Weng:2018mmf}. The masses are all in units of MeV.}
\label{fig:cccccq+bbbbbq}
\end{figure*}

First we consider the $cccccq$ dibaryons.
In Fig.~\ref{fig:cccccq+bbbbbq}(a--b), We plot the relative position of these dibaryons, along with the corresponding baryon-baryon thresholds which they can decay into through quark rearrangement.
From the figure, we can easily see that all states are above thresholds.
They may all be \emph{broad} states since they can decay into two baryons through $S$-wave~\cite{Jaffe:1976ig}.
We find that the quantum number of the ground states are $J^{P}=1^{+}$ in both cases.
They are the $D(cccccn,8516,1^{+})$ and the $D(cccccs,8617,1^{+})$ states, respectively.
Unlike the conventional mesons and baryons, the $cccccq$ scalar states, which have lower spin, are heavier then the axial-vector states by almost 100~MeV.
Similar to the fully heavy $c^{5}b$ and $b^{5}c$ dibaryons, the underlying reason of this mass reversal is that the Pauli principle imposes large restriction over the wave
functions~\cite{Weng:2022ohh}.
The colorelectric interaction
\begin{equation}\label{eqn:Hcolor:cccccq}
\Braket{H_{\text{C}}\left(cccccq\right)}
=
2m_{cc}+m_{cq}\,,
\end{equation}
is identical for the two states.
By introducing the $\mathrm{SU}(6)_{cs}=\mathrm{SU}(3)_{c}\otimes\mathrm{SU}(2)_{s}$ group, the chromomagnetic interaction can be written as~\cite{Jaffe:1976ih,Weng:2022ohh}
\begin{equation}\label{eqn:HCM:cccccq}
\Braket{H_{\text{CM}}\left(cccccq\right)}
=
2v_{cc}+v_{cq}\left[1+\frac{S\left(S+1\right)}{12}-\frac{\Braket{\text{C}_{6}\left(cccccq\right)}}{32}\right]\,,
\end{equation}
where $\text{C}_{6}$ is the $\mathrm{SU}(6)_{cs}$ Casimir operator.
The spin term does contribute larger mass for the higher spin state, but the $\text{C}_{6}$ term has opposite effect.
In the $\mathrm{SU}(6)_{cs}$ group, the Pauli principle requires that the five charm quarks be anti-symmetric
\begin{equation}\label{eqn:rep:SU6:ccccc}
\Yvcentermath1
\young(c,c,c,c,c)_{cs}
\sim
\young(cc,cc,c)_{c}
\otimes
\young(ccc,cc)_{s}\,.
\end{equation}
Adding the light quark, we have two possible $\text{SU}(6)_{cs}$ representations.
Namely a $J=0$ representation
\begin{equation}\label{eqn:rep:SU6:cccccq:S0}
\Yvcentermath1
\young(c,c,c,c,c,q)_{cs}
\sim
\young(cc,cc,cq)_{c}
\otimes
\young(ccc,ccq)_{s}\,,
\end{equation}
and a $J=1$ representation
\begin{equation}\label{eqn:rep:SU6:cccccq:S1}
\Yvcentermath1
\young(cq,c,c,c,c)_{cs}
\sim
\young(cc,cc,cq)_{c}
\otimes
\young(cccq,cc)_{s}\,,
\end{equation}
where we have require the states to be of color-singlet.
The former representation ($J=0$) is a $\text{SU}(6)_{cs}$ singlet with $\braket{\text{C}_{6}}=0$, while the latter one ($J=1$) is a 35-plet with $\braket{\text{C}_{6}}=48$~\cite{Jaffe:1976ih}.
Thus we have
\begin{equation}\label{eqn:HCM:cccccq:2}
\Braket{H_{\text{CM}}\left(cccccq\right)}
=
\left\{
\begin{split}
&2v_{cc}+v_{cq}~\text{ for }~J=0\,,\\
&2v_{cc}-\frac{v_{cq}}{3}~\text{ for }~J=1\,.
\end{split}
\right.
\end{equation}
The effect of $\text{C}_{6}$ term surpasses that of the spin term.
Thus the chromomagnetic interaction favors the axial-vector state over the scalar state.

The spectra of their bottom partners, $bbbbbq$ dibaryons, share similar pattern.
As shown in Fig.~\ref{fig:cccccq+bbbbbq}(c--d), the ground states are the $D(bbbbbn,24526,1^{+})$ and the $D(bbbbbs,24612,1^{+})$, respectively.
While the excitations are the $D(bbbbbn,24553,0^{+})$ and the $D(bbbbbs,24624,0^{+})$ states, respectively.
The corresponding mass splittings are about $20~\text{MeV}$, which is much smaller than that of their charm partners.
The reason is that the splitting is inversely proportional to the quark masses [see Eq.~\eqref{eqn:HCM:cccccq:2}]
\begin{equation}
\Delta{m}\left(Q^{5}q\right){\propto}v_{Qq}{\propto}1/m_{Q}m_{q}\,.
\end{equation}
The bottom quark is much heavier than the charm quark, thus the mass splitting of the $bbbbbq$ dibaryons is smaller than that of the $cccccq$ dibaryons.

Besides the masses, the eigenvectors can also be used to investigate the decay properties of the dibaryons.
We can transform the wave function into the $q^{3}{\otimes}q^{3}$ configuration.
The $q^{3}$ component can be either of color-singlet or of color-octet.
The $\ket{(qqq)^{1_{c}}{\otimes}(qqq)^{1_{c}}}$ dibaryon can easily dissociate into two baryons in relative $S$-wave.
We call this kind of decay the ``Okubo-Zweig-Iizuka- (OZI-)superallowed'' decay.
On the other hand, the $\ket{(qqq)^{8_{c}}{\otimes}(qqq)^{8_{c}}}$ dibaryon cannot fall apart without the gluon exchange.
For simplicity, we follow Refs.~\cite{Jaffe:1976ig,Jaffe:1976ih,Strottman:1979qu,Weng:2022ohh} and only consider the ``OZI-superallowed'' decays in this work.
In Table~\ref{table:cccccq+bbbbbq:decay:eigenvector}, we transform the eigenvectors of the $cccccq$ ($bbbbbq$) dibaryons into the $ccc{\otimes}ccq$ ($bbb{\otimes}bbq$) configuration.
For simplicity, we only present the color-singlet components, and rewrite the bases as a direct product of two baryons
\begin{equation}
\Psi
=
\sum_{i}c_{i}\Ket{\psi_{i}\left(B{\otimes}B\right)}
+\cdots\,,
\end{equation}
where $c_{i}$ represents the overlap between the dibaryon and a particular baryon~$\times$~baryon channel.
For each decay channel, the decay width is proportional to the square of the coefficient $c_{i}$ of the corresponding component in the eigenvector, and also depends on the phase space.
For two body $L$-wave decay, the partial decay width reads~\cite{Gao-1992-Group,Weng:2019ynv}
\begin{equation}\label{eqn:width}
\Gamma_{i}=\gamma_{i}\alpha\frac{k^{2L+1}}{m^{2L}}{\cdot}|c_i|^2,
\end{equation}
where $\gamma_{i}$ is a quantity related to the decay dynamics, $\alpha$ is an effective coupling constant, $k$ is the momentum of the final baryons in the rest frame of the initial state, $m$ is the dibaryon mass, and $L$ is the relative partial wave between the two baryons.
For the decay processes studied in this work, $(k/m)^2$'s are all of $\mathcal{O}(10^{-2})$ or even smaller.
The higher wave decays are all suppressed.
Thus we only consider the $S$-wave decays.
%
%
Generally, $\gamma_{i}$ depends on the spatial wave functions of the initial and final states, which are different for each decay process.
In the quark model, the spatial wave functions of the ground state $1/2^{+}$ and $3/2^{+}$ baryons  are similar.
Thus for \emph{each} dibaryon state, we have
\begin{equation}
\gamma_{B_{1}B_{2}}
\approx
\gamma_{B_{1}B_{2}^{*}}
\approx
\gamma_{B_{1}^{*}B_{2}}
\approx
\gamma_{B_{1}^{*}B_{2}^{*}}\,,
\end{equation}
where $B_{i}$ and $B_{i}^{*}$ are the $1/2^{+}$ and $3/2^{+}$ baryons with the same flavor contents.
In Table~\ref{table:cccccq+bbbbbq:decay:kci2}, we calculate the values of $k\cdot|c_i|^2$ for the $cccccq$ and $bbbbbq$ dibaryons.
Then we compare their relative widths in Table~\ref{table:cccccq+bbbbbq:decay:Ratio}.

For the $cccccq$ and $bbbbbq$ dibaryons, the scalar states only have one $S$-wave decay channel consists of two $3/2^{+}$ baryons.
They are broad states since their phase spaces are quite large (see Fig.~\ref{fig:cccccq+bbbbbq}).
The axial-vector states have two $S$-wave decay channels.
From Table~\ref{table:cccccq+bbbbbq:decay:Ratio}, we have
\begin{equation}
\frac
{\Gamma\left[D(cccccn,8516,1^{+}){\to}\Omega_{ccc}^{*}\Xi_{cc}\right]}
{\Gamma\left[D(cccccn,8516,1^{+}){\to}\Omega_{ccc}^{*}\Xi_{cc}^{*}\right]}
\sim
1.3\,,
\end{equation}
\begin{equation}
\frac
{\Gamma\left[D(cccccs,8617,1^{+}){\to}\Omega_{ccc}^{*}\Omega_{cc}\right]}
{\Gamma\left[D(cccccs,8617,1^{+}){\to}\Omega_{ccc}^{*}\Omega_{cc}^{*}\right]}
\sim
1.5\,,
\end{equation}
\begin{equation}
\frac
{\Gamma\left[D(bbbbbn,24526,1^{+}){\to}\Omega_{bbb}^{*}\Xi_{bb}\right]}
{\Gamma\left[D(bbbbbn,24526,1^{+}){\to}\Omega_{bbb}^{*}\Xi_{bb}^{*}\right]}
\sim
1.0\,,
\end{equation}
and
\begin{equation}
\frac
{\Gamma\left[D(bbbbbs,24612,1^{+}){\to}\Omega_{bbb}^{*}\Omega_{bb}\right]}
{\Gamma\left[D(bbbbbs,24612,1^{+}){\to}\Omega_{bbb}^{*}\Omega_{bb}^{*}\right]}
\sim
0.9\,.
\end{equation}
For axial-vector states, the partial decay widths of the two channels are compatible.
%

\subsection{The $ccccbq$ and $bbbbcq$ systems}

\begin{figure*}
\begin{tabular}{ccc}
\includegraphics[width=500pt]{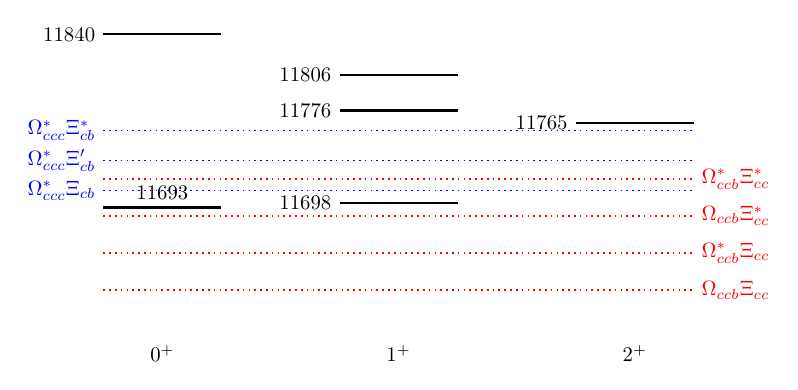}\\
(a) $ccccbn$ states\\
&&\\
\includegraphics[width=500pt]{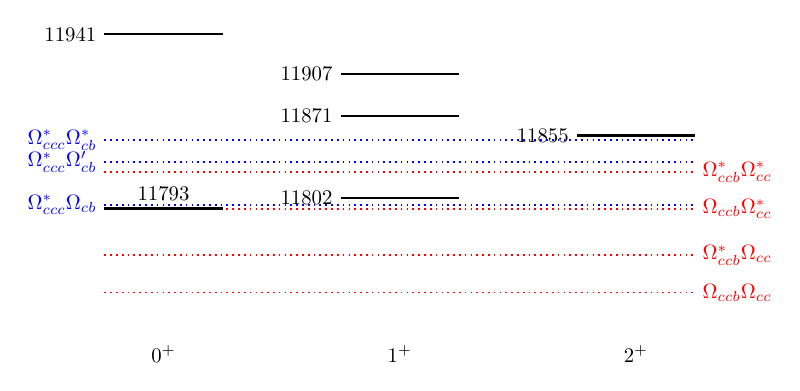}\\
(b) $ccccbs$ states\\
\end{tabular}
\caption{Masses of the $ccccbq$ dibaryons. The dotted lines indicate various baryon-baryon thresholds, where the baryon masses are calculated in the same model~\cite{Weng:2018mmf}. The masses are all in units of MeV.}
\label{fig:ccccbq}
\end{figure*}

\begin{figure*}
\begin{tabular}{ccc}
\includegraphics[width=500pt]{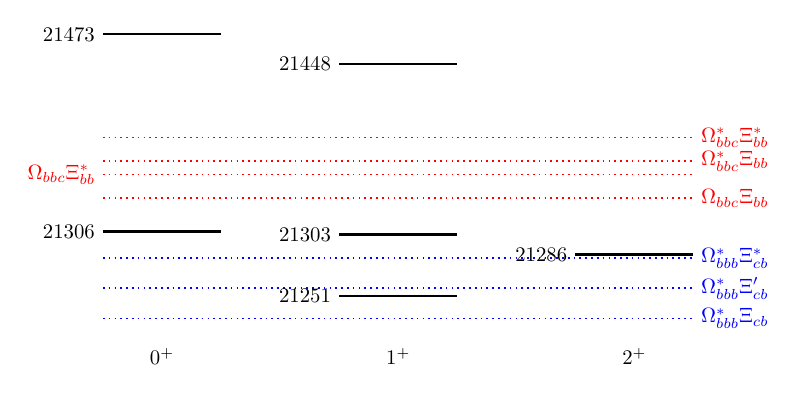}\\
(a) $bbbbcn$ states\\
&&\\
\includegraphics[width=500pt]{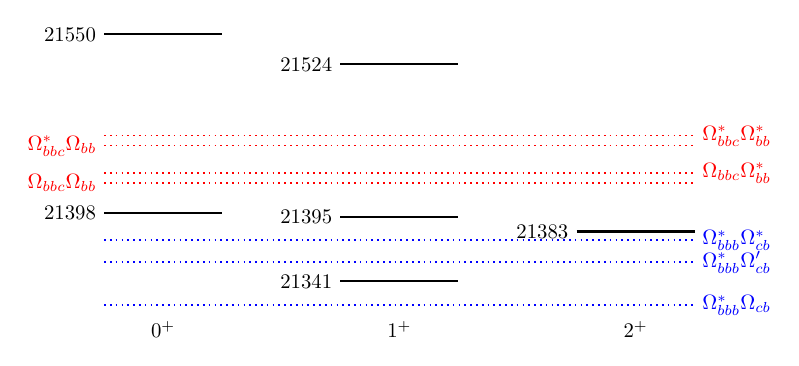}\\
(b) $bbbbcs$ states\\
\end{tabular}
\caption{Masses of the $bbbbcq$ dibaryons. The dotted lines indicate various baryon-baryon thresholds, where the baryon masses are calculated in the same model~\cite{Weng:2018mmf}. The masses are all in units of MeV.}
\label{fig:bbbbcq}
\end{figure*}

Now we turn to the $ccccbq$ and $bbbbcq$ dibaryons.
We list their masses and eigenvectors in Table~\ref{table:mass:cccccq+bbbbbq+ccccbq+bbbbcq} and plot their relative position in Fig.~\ref{fig:ccccbq}--\ref{fig:bbbbcq}, along with the possible decay channels.
Similar to the $cccccq$ and $bbbbbq$ dibaryon cases, all $ccccbq$ dibaryon states are above baryon-baryon thresholds.
The ground states are both scalar states.
They are $D(ccccbn,11693,0^{+})$ and $D(ccccbs,11793,0^{+})$, respectively.
The heaviest states are also scalar states.
Namely $D(ccccbn,11840,0^{+})$ and $D(ccccbs,11941,0^{+})$.
The mass splittings of the $ccccbn$ and $ccccbs$ dibaryons are $247~\text{MeV}$ and $248~\text{MeV}$, respectively.

Besides the masses, the eigenvectors also provide important information of the dibaryons.
As shown in Tables~\ref{table:wavefunc:color+spin:12x3x4x5x6:12x3x45x6}--\ref{table:wavefunc:total}, there are two scalar bases, three axial-vector bases and one tensor base.
Their color configurations can be of $\ket{(c_{1}c_{2}c_{3}c_{4})^{\bar{6}_{c}}(b_{5}q_{6})^{6_{c}}}$ or $\ket{(c_{1}c_{2}c_{3}c_{4})^{3_{c}}(b_{5}q_{6})^{\bar{3}_{c}}}$.
For simplicity, we denote them as $\bar{6}_{c}{\otimes}6_{c}$ and $3_{c}{\otimes}\bar{3}_{c}$.
From Table~\ref{table:mass:cccccq+bbbbbq+ccccbq+bbbbcq}, we find that the ground states of the $ccccbn$ and $ccccbs$ dibaryons are both dominated by the color-sextet components.
More precisely, the $D(ccccbn,11693,0^{+})$ has $80.9\%$ of the $\bar{6}_{c}{\otimes}6_{c}$ component and the $D(ccccbs,11793,0^{+})$ has $76.6\%$.
Moreover, the lowest axial-vector states are also dominated by the color-sextet components.
The $D(ccccbn,11698,1^{+})$ and $D(ccccbs,11802,1^{+})$ possess $81.4\%$ and $76.0\%$ of the $\bar{6}_{c}{\otimes}6_{c}$ component, respectively.
On the other hand, all higher states are dominated by the color-triplet components.
The underlying reason is the color interaction.
More precisely, we have
\begin{equation}\label{eqn:HCM:ccccbq}
\Braket{H_{\text{C}}\left(ccccbq\right)}
=
2m_{cc}+m_{bq}
-
\frac{3}{8}\delta{m}\cdot\mathrm{C}_{3}\left(cccc\right)
\end{equation}
where $\text{C}_{3}$ is the $\mathrm{SU}(3)_{c}$ Casimir operator and
\begin{equation}
\delta{m}
=
\frac{m_{cc}+m_{bq}-m_{cb}-m_{cq}}{4}\,.
\end{equation}
Inserting the parameters in Table~\ref{table:parameter:qq}, we have
\begin{equation}
\left\{
\begin{split}
\delta{m}\left(ccccbn\right)=21.93~\text{MeV}\,,\\
\delta{m}\left(ccccbs\right)=16.64~\text{MeV}\,.
\end{split}
\right.
\end{equation}
Note that the $\mathrm{C}_{3}(cccc)$ is diagonal in the $\bar{6}_{c}{\otimes}6_{c}$ and $3_{c}{\otimes}\bar{3}_{c}$ color bases, with matrix elements $40/3$ and $16/3$, respectively.
Thus
\begin{equation}
\Braket{H_{\text{C}}\left(ccccbq\right)}
=
\left\{
\begin{split}
2m_{cc}+m_{bq}-5\delta{m}~\text{for}~\bar{6}_{c}{\otimes}6_{c}\,,\\
2m_{cc}+m_{bq}-2\delta{m}~\text{for}~3_{c}{\otimes}\bar{3}_{c}\,.
\end{split}
\right.
\end{equation}
The color interaction favors the $\bar{6}_{c}{\otimes}6_{c}$ configuration.
Then we obtain the two-band structure in the mass spectrum.
The CM interaction is suppressed by $1/m_{Q}$ and thus cannot alter the order.
Alas, the CM interaction mixes the states and gives small shift of the bands.
%

%
Unlike the $ccccbq$ dibaryons, the quantum numbers of the $bbbbcq$ ground states become $J^{P}=1^{+}$.
All $bbbbcq$ dibaryons are above their baryon-baryon threshold.
However, some of them may be relatively narrow because they are not too much higher than their $S$-wave decay channels (see Fig.~\ref{fig:bbbbcq}).
The $D(bbbbcn,21251,1^{+})$ can decay to the $\Omega_{bbb}^{*}\Xi_{cb}$ mode, but their mass difference is only $19~\text{MeV}$.
The $D(bbbbcn,21306,0^{+})$ has three decay channels.
It can only decay into the $\Omega_{bbb}^{*}\Xi_{cb}$ and $\Omega_{bbb}^{*}\Xi_{cb}'$ channels by $D$-wave due to the conservation of the angular momentum.
According to Eq.~\eqref{eqn:width}, these channels are highly suppressed.
On the other hand, it can also decay into the $\Omega_{bbb}^{*}\Xi_{cb}^{*}$ channel through $S$-wave, but their mass difference is only $23~\text{MeV}$.
The $bbbbcs$ dibaryons are similar, the $D(bbbbcs,21341,1^{+})$/$D(bbbbcs,21398,0^{+})$ is only $21/23~\text{MeV}$ above the $\Omega_{bbb}^{*}\Omega_{cb}$/$\Omega_{bbb}^{*}\Omega_{cb}^{*}$ channel.
Due to their small phases, these stats are expected to be relatively narrow.

Another difference of the $ccccbq$ and $bbbbcq$ systems is that the dominant color configuration of the lower mass $bbbbcq$ dibaryon states becomes $3_{c}{\otimes}\bar{3}_{c}$.
For instance, the $D(bbbbcn,21251,1^{+})$ has $99.2\%$ of $3_{c}{\otimes}\bar{3}_{c}$ component, and the $D(bbbbcs,21341,1^{+})$ has $99.5\%$.
We again resort to the color interaction
\begin{equation}\label{eqn:HCM:bbbbcq}
\Braket{H_{\text{C}}\left(bbbbcq\right)}
=
2m_{bb}+m_{cq}
-
\frac{3}{8}\delta{m}'\cdot\mathrm{C}_{3}\left(bbbb\right)
\end{equation}
where
\begin{equation}
\delta{m}'
=
\frac{m_{bb}+m_{cq}-m_{cb}-m_{bq}}{4}\,.
\end{equation}
The main difference between $ccccbq$ and $bbbbcq$ dibaryons is that the $\delta{m}$'s are positive, while the $\delta{m}'$'s are negative
\begin{equation}
\left\{
\begin{split}
\delta{m}'\left(bbbbcn\right)=-54.70~\text{MeV}\,,\\
\delta{m}'\left(bbbbcs\right)=-49.40~\text{MeV}\,.
\end{split}
\right.
\end{equation}
Thus the lower mass states are dominated by the $3_{c}{\otimes}\bar{3}_{c}$ components, while the higher one have more $\bar{6}_{c}{\otimes}6_{c}$ components.

Before concluding this section, we would like to compare the present results with that of Refs.~\cite{Weng:2020jao,Weng:2021hje,Weng:2021ngd}.
It is interesting to note that $ccccbq$ dibaryons are very similar to the triply heavy $\bar{c}\bar{c}bq$ tetraquarks and the fully heavy tetraquarks like $\bar{b}\bar{b}bc$.
On the other hand, the $bbbbcq$ dibaryons are more like the triply heavy $\bar{b}\bar{b}cq$ tetraquarks and the doubly heavy tetraquarks.
The underlying reason is the similarity between the $cccc$/$bbbb$ cluster in the dibaryons and the $\bar{c}\bar{c}$/$\bar{b}\bar{b}$ cluster in the tetraquarks.
For example, their possible color representations are both $\{\bar{6}_{c},3_{c}\}$, while their possible spins are both $\{0,1\}$ as well.
%
In fact, the similarity can also be inferred from the color interaction (taking $\bar{c}\bar{c}bq$ as an example)
\begin{equation}\label{eqn:HCM:ccbq}
\Braket{H_{\text{C}}\left(\bar{c}\bar{c}bq\right)}
=
m_{cc}+m_{bq}
-
\frac{3}{8}\delta\tilde{m}\cdot\mathrm{C}_{3}\left(\bar{c}\bar{c}\right)
\end{equation}
where
\begin{equation}
\delta\tilde{m}
=
\frac{m_{cc}+m_{bq}-m_{\bar{c}b}-m_{\bar{c}q}}{4}\,.
\end{equation}
A more detailed dynamical study of the dependence of the spectrum and wave function with respect to the quark masses would be very important for decoding the nature of dibaryons.

Next we consider their decay properties.
We transform the eigenvectors of the $ccccbq$ ($bbbbcq$) dibaryons into the $ccc{\otimes}cbq$ ($bbb{\otimes}cbq$) and $ccb{\otimes}ccq$ ($bbc{\otimes}bbq$) configurations.
For all possible channels, we calculate the values of $k{\cdot}|c_{i}|^{2}$ and the ratios of the partial decay widths.
Note that for a dibaryon state decays into two channels through two different quark rearrangements, the $\gamma_{i}$'s, which depend on the wave functions of final states, may not equal (or approximately equal).
Thus we compare only the decay width ratios between channels through same quark rearrangement type.
the corresponding results are listed in Table~\ref{table:ccccbq+bbbbcq:decay:eigenvector}--\ref{table:ccccbq+bbbbcq:decay:Ratio}.
It is easy to see that the decay properties of the $ccccbq$ and the $bbbbcq$ dibaryons are quite different.
Taking the higher scalar states as example.
For the $ccccbq$ dibaryons, we have
\begin{equation}
\frac{\Gamma\left[D(ccccbn,11840,0^{+}){\to}\Omega_{ccb}^{*}\Xi_{cc}^{*}\right]}{\Gamma\left[D(ccccbn,11840,0^{+}){\to}\Omega_{ccb}\Xi_{cc}\right]}
\sim
11.0\,,
\end{equation}
and
\begin{equation}
\frac{\Gamma\left[D(ccccbs,11941,0^{+}){\to}\Omega_{ccb}^{*}\Omega_{cc}^{*}\right]}{\Gamma\left[D(ccccbs,11941,0^{+}){\to}\Omega_{ccb}\Omega_{cc}\right]}
\sim
19.5\,.
\end{equation}
The former modes always dominate.
But for the $bbbbcq$ dibaryons, we have
\begin{equation}
\frac{\Gamma\left[D(bbbbcn,21473,0^{+}){\to}\Omega_{bbc}^{*}\Xi_{bb}^{*}\right]}{\Gamma\left[D(bbbbcn,21473,0^{+}){\to}\Omega_{bbc}\Xi_{bb}\right]}
\sim
2.6\,,
\end{equation}
and
\begin{equation}
\frac{\Gamma\left[D(bbbbcs,21550,0^{+}){\to}\Omega_{bbc}^{*}\Omega_{bb}^{*}\right]}{\Gamma\left[D(bbbbcs,21550,0^{+}){\to}\Omega_{bbc}\Omega_{bb}\right]}
\sim
2.5\,.
\end{equation}
The two modes are compatible.
As discussed above, the wave functions of the $ccccbq$ and the $bbbbcq$ dibaryons are quite different,
which results in the difference of their decay properties.
%

\subsection{The $cccbbq$ and $bbbccq$ systems}

The eigenvalues and eigenvectors of the $cccbbq$ system are listed in Table~\ref{table:mass:cccbbq}.
We transform their eigenvectors into the $ccc{\otimes}bbq$, $ccb{\otimes}cbq$ and $ccq{\otimes}bbc$ configurations, as shown in Table~\ref{table:cccbbq:decay:eigenvector}.
From these tables, we see that some eigenstates couple very strongly with two $S$-wave baryons.
For example, the second scalar state
\begin{equation}
D\left(cccbbn,14981,0^{+}\right)
=
0.899
\Omega_{ccc}^{*}\Xi_{bb}^{*}
+
\cdots\,.
\end{equation}
This state couples strongly $(80.8\%)$ to the $\Omega_{ccc}^{*}\Xi_{bb}^{*}$ scattering state.
It might be a broad state which belongs to the continuum.
Various calculations also found existence of such kind of states in tetraquarks~\cite{Hogaasen:2005jv,Cui:2006mp,Weng:2020jao,Weng:2021ngd,Guo:2022crh}, pentaquarks~\cite{Weng:2019ynv} and dibaryons~\cite{Alcaraz-Pelegrina:2022fsi,Weng:2022ohh}.
Besides the scalar case, the $cccbbn$ states of $14984~\text{MeV}$ (with $J^{P}=1^{+}$), $14976~\text{MeV}$ (with $J^{P}=2^{+}$) and $14974~\text{MeV}$ (with $J^{P}=3^{+}$) also couple strongly to the $\Omega_{ccc}^{*}$ and $\Xi_{bb}^{*}$ baryons, while the states of $14953~\text{MeV}$ (with $J^{P}=1^{+}$) and $14962~\text{MeV}$ (with $J^{P}=2^{+}$) couple very strongly to the $\Omega_{ccc}^{*}{\otimes}\Xi_{bb}$ channel.
They may be scattering states.
On the other hand, the $cccbbs$ state of $15053~\text{MeV}$ with $J^{P}=3^{+}$ is a $\Omega_{ccc}^{*}{\otimes}\Omega_{bb}$ scattering state.
Among them, $D(cccbbn,14984,1^{+})$ state couples relatively weaker ($73.8\%$) to the $\Omega_{ccc}^{*}{\otimes}\Xi_{bb}^{*}$ channel, thus we cannot rule out the possibility that it is a dibaryons.
To draw a definite conclusion, dynamical studies like the complex scaling methods~\cite{Myo:2020rni} are needed, which is beyond the present work.
For clarity, we indicate the scattering states in the last column in Table~\ref{table:mass:cccbbq}.
Going to the $bbbccq$ dibaryons, we also find various scattering states.
They are listed in the fifth column of Table~\ref{table:mass:bbbccq}.

\begin{figure*}
\begin{tabular}{ccc}
\includegraphics[width=450pt]{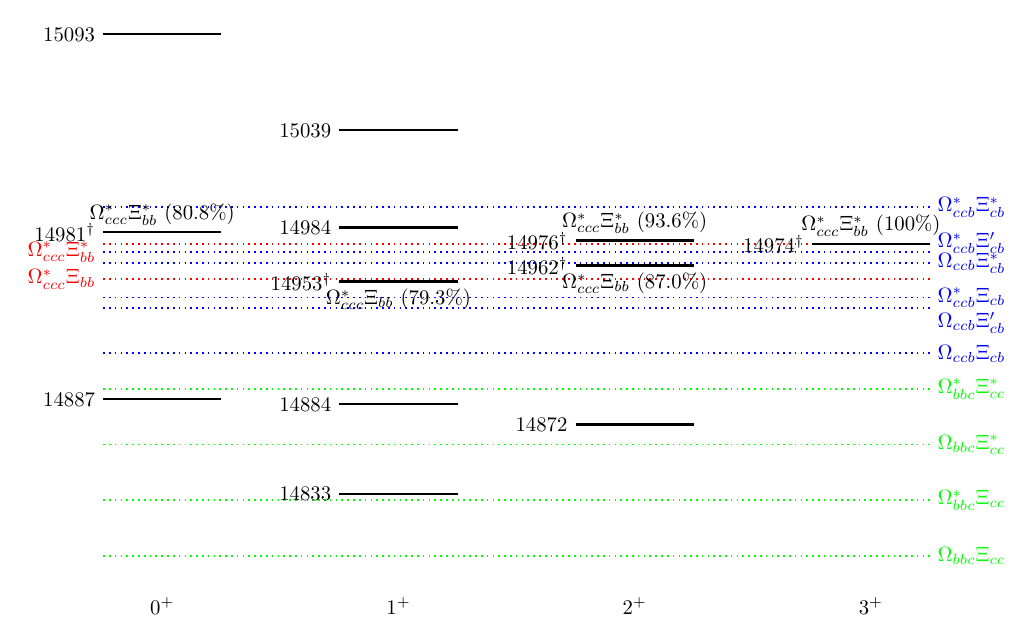}\\
(a) $cccbbn$ states\\
&&\\
\includegraphics[width=450pt]{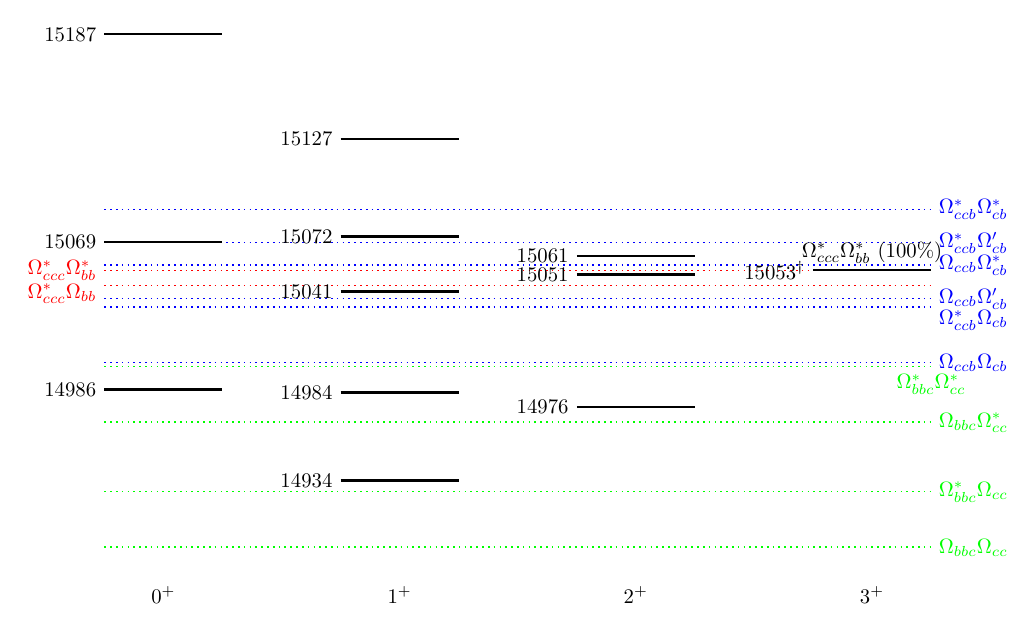}\\
(b) $cccbbs$ states\\
\end{tabular}
\caption{Masses of the $cccbbq$ dibaryons. The dotted lines indicate various baryon-baryon thresholds, where the baryon masses are calculated in the same model~\cite{Weng:2018mmf}. Possible scattering states are marked with a dagger ($\dagger$), along with the proportion of their dominant components. The masses are all in units of MeV.}
\label{fig:cccbbq}
\end{figure*}

\begin{figure*}
\begin{tabular}{ccc}
\includegraphics[width=450pt]{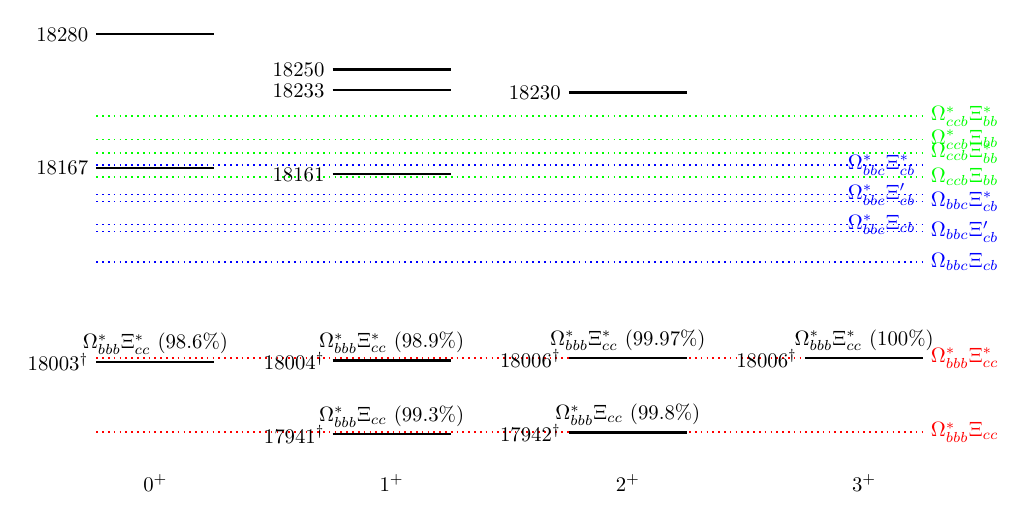}\\
(a) $bbbccn$ states\\
&&\\
\includegraphics[width=450pt]{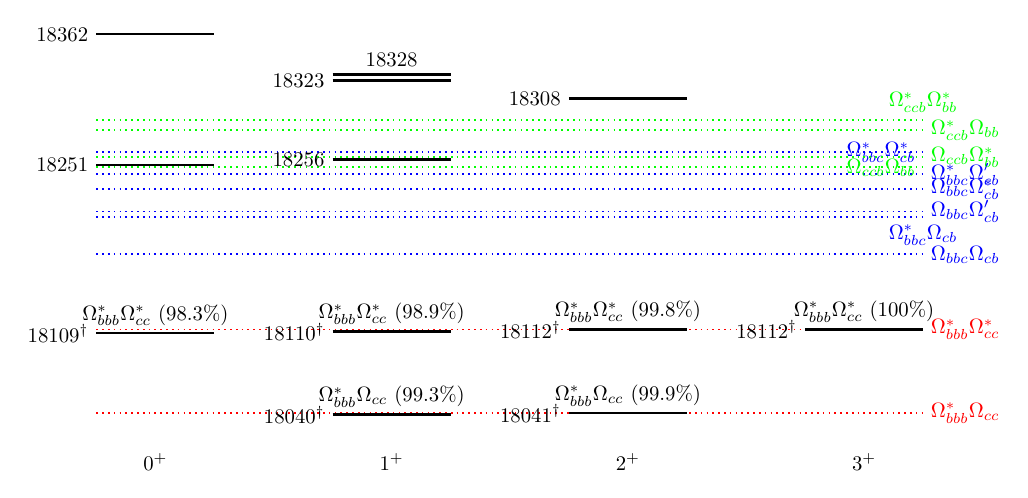}\\
(b) $bbbccs$ states\\
\end{tabular}
\caption{Masses of the $bbbccq$ dibaryons. The dotted lines indicate various baryon-baryon thresholds, where the baryon masses are calculated in the same model~\cite{Weng:2018mmf}. Possible scattering states are marked with a dagger ($\dagger$), along with the proportion of their dominant components. The masses are all in units of MeV.}
\label{fig:bbbccq}
\end{figure*}

We plot the relative position of the $cccbbq$ and $bbbccq$ dibaryons in Figs.~\ref{fig:cccbbq}--\ref{fig:bbbccq}.
For comparison, we also plot the possible scattering states.
They are marked with a dagger ($\dagger$), along with the proportion of their dominant components.
We can easily see that they all lie close to the corresponding $QQQ$ baryon-$QQ'q$ baryon thresholds.
Interestingly, we find that the scattering states of the $cccbbq$ systems lie in the middle of the spectra, whit those of the $bbbccq$ systems lie in the bottom of the spectra.
The reason is as follows.
From Tables~\ref{table:wavefunc:color+spin:12x3x4x5x6:12x3x45x6}--\ref{table:wavefunc:total}, we see that the $Q^{3}Q'^{2}q$ dibaryons has three possible color bases
\begin{align}
&8_{c}{\otimes}6_{c}{\otimes}3_{c}:
\left\{\left[\left(QQQ\right)^{8_{c}}{\otimes}\left(Q'Q'\right)^{6_{c}}\right]^{\bar{3}_{c}}{\otimes}q^{3_{c}}\right\}^{1_{c}}\,,\\
&8_{c}{\otimes}\bar{3}_{c}{\otimes}3_{c}:
\left\{\left[\left(QQQ\right)^{8_{c}}{\otimes}\left(Q'Q'\right)^{\bar{3}_{c}}\right]^{\bar{3}_{c}}{\otimes}q^{3_{c}}\right\}^{1_{c}}\,,\\
&1_{c}{\otimes}\bar{3}_{c}{\otimes}3_{c}:
\left\{\left[\left(QQQ\right)^{1_{c}}{\otimes}\left(Q'Q'\right)^{\bar{3}_{c}}\right]^{\bar{3}_{c}}{\otimes}q^{3_{c}}\right\}^{1_{c}}\,.
\end{align}
Note that the last one is the dominant component of the scattering states.
For these systems, the color interaction reads
\begin{align}
&\Braket{H_{\text{C}}\left(Q^{3}Q'^{2}q\right)}
\notag\\
={}&
M
-\frac{3}{8}\delta{m}_{\alpha}\cdot\mathrm{C}_{3}\left(QQQ\right)
-\frac{3}{8}\delta{m}_{\beta}\cdot\mathrm{C}_{3}\left(Q'Q'\right)\,,
\end{align}
where
\begin{equation}
M=
\frac{3m_{QQ}+2m_{Q'Q'}+m_{qQ}+m_{qQ'}-m_{QQ'}}{2}\,,
\end{equation}
\begin{equation}
\delta{m}_{\alpha}
=
\frac{m_{QQ}+m_{qQ'}-m_{qQ}-m_{QQ'}}{4}\,,
\end{equation}
and
\begin{equation}
\delta{m}_{\beta}
=
\frac{m_{Q'Q'}+m_{qQ}-m_{qQ'}-m_{QQ'}}{4}\,.
\end{equation}
The color interaction is diagonal over the three bases.
For the $cccbbq$ dibaryons, $\delta{m}_{\alpha}\sim20~\text{MeV}$ and $\delta{m}_{\beta}\sim-50~\text{MeV}$.
Thus the $\mathrm{C}_{3}(ccc)$ term favors the bases with $(ccc)^{8_{c}}$ cluster, while the $\mathrm{C}_{3}(bb)$ term favors the bases with $(bb)^{\bar{3}_{c}}$ cluster.
Since $|\delta{m}_{\alpha}|<|\delta{m}_{\beta}|$, the $\mathrm{C}_{3}(bb)$ term dominates the mass splitting while the $\mathrm{C}_{3}(ccc)$ term give smaller splittings afterward.
In summary, we have
\begin{equation}
8_{c}{\otimes}\bar{3}_{c}{\otimes}3_{c}
<
1_{c}{\otimes}\bar{3}_{c}{\otimes}3_{c}
\ll
8_{c}{\otimes}6_{c}{\otimes}3_{c}\,.
\end{equation}
Thus the scattering states of the $cccbbq$ systems lie in the middle of the spectra.
For the $bbbccq$ dibaryons, $\delta{m}_{\alpha}\sim-50~\text{MeV}$ and $\delta{m}_{\beta}\sim20~\text{MeV}$.
With similar argument, we have
\begin{equation}
1_{c}{\otimes}\bar{3}_{c}{\otimes}3_{c}
\ll
8_{c}{\otimes}6_{c}{\otimes}3_{c}
<
8_{c}{\otimes}\bar{3}_{c}{\otimes}3_{c}\,.
\end{equation}
Thus the scattering states of the $bbbccq$ systems lie in the bottom of the spectra.
When including the CM interaction, these color bases will be mixed.
Since the splittings between $1_{c}{\otimes}\bar{3}_{c}{\otimes}3_{c}$ base and other bases are quite large in the $bbbccq$ system, the corresponding mixing effects become small.
Thus the proportions of $bbbccq$ scattering states' dominant components are always very large ($\sim100\%$).
On the contrary, the splitting between the $1_{c}{\otimes}\bar{3}_{c}{\otimes}3_{c}$ and $8_{c}{\otimes}\bar{3}_{c}{\otimes}3_{c}$ bases is quite small in the $cccbbq$ system, the corresponding mixing effect becomes pretty important.
So the proportions of $cccbbq$ scattering states' dominant components become significantly smaller (see Table~\ref{table:mass:cccbbq}).

After identifying the scattering states, the other states are genuine dibaryons.
They all lie far above their $S$-wave decay channel(s).
Thus they may be broad states.
We also study their decay properties, which can be found in Tables~\ref{table:cccbbq:decay:eigenvector}--\ref{table:bbbccq:decay:Ratio}.
%

\section{Conclusions}
\label{Sec:Conclusion}

In this work, we have systematically studied the mass spectrum of dibaryons with five heavy quarks, in the framework of an extended chromomagnetic model, which includes both the colorelectric and chromomagnetic interactions.
To do this, we first constructed the flavor~$\otimes$~color~$\otimes$~spin wave functions of the $Q^{5}q$ dibaryons to satisfy the $\mathrm{SU}(2)_{f}$ isospin symmetry and the Pauli principle.
Then we systematically calculated the interaction Hamiltonian under these bases and obtain the corresponding masses.
The numerical results suggest that all states are above the corresponding baryon-baryon thresholds.
Most of them are expected to be broad states.
However, the $D(bbbbcn,21306,0^{+})$, $D(bbbbcn,21251,1^{+})$, $D(bbbbcs,21398,0^{+})$ and $D(bbbbcs,21341,1^{+})$ both have only one $S$-wave decay channel and they lie above the corresponding channels by only about $20~\text{MeV}$.
Due to the small phases, these states may be relatively narrow.
In addition to computing the masses, we also use the wave functions to investigate the decay properties of the dibaryons.
We hope that our studies can be of help for future experimental searches.
%

\section*{Acknowledgments}

X.Z.W is grateful to Professor~Marek~Karliner and Professor~Shi-Lin~Zhu for helpful comments and discussions.
%
This project was supported by the NSFC-ISF under Grant No. 3423/19.
%




%

\begin{table*}
\centering
\caption{The eigenvectors of the $cccccq$ and $bbbbbq$ dibaryon states. The masses are all in units of MeV.}
\label{table:cccccq+bbbbbq:decay:eigenvector}
\begin{tabular}{ccccccccccc}
\toprule[1pt]
\toprule[1pt]
System&$J^{P}$&Mass
&$\Omega_{ccc}^{*}\Xi_{cc}^{*}$
&$\Omega_{ccc}^{*}\Xi_{cc}$
&
System&$J^{P}$&Mass
&$\Omega_{bbb}^{*}\Xi_{bb}^{*}$
&$\Omega_{bbb}^{*}\Xi_{bb}$\\
\midrule[1pt]
$cccccn$&$0^{+}$
&$8599$&$-0.447$&
&
$bbbbbn$&$0^{+}$
&$24553$&$-0.447$\\
&$1^{+}$
&$8516$&$-0.333$&$0.298$
&
&$1^{+}$
&$24526$&$-0.333$&$0.298$\\
\midrule[1pt]
System&$J^{P}$&Mass
&$\Omega_{ccc}^{*}\Omega_{cc}^{*}$
&$\Omega_{ccc}^{*}\Omega_{cc}$
&
System&$J^{P}$&Mass
&$\Omega_{bbb}^{*}\Omega_{bb}^{*}$
&$\Omega_{bbb}^{*}\Omega_{bb}$\\
\midrule[1pt]
$cccccs$&$0^{+}$
&$8711$&$-0.447$&
&
$bbbbbs$&$0^{+}$
&$24624$&$-0.447$\\
&$1^{+}$
&$8617$&$-0.333$&$0.298$
&
&$1^{+}$
&$24612$&$-0.333$&$0.298$\\
\bottomrule[1pt]
\bottomrule[1pt]
\end{tabular}
\end{table*}
%
\begin{table*}
\centering
\caption{The values of $k\cdot|c_{i}|^2$ for the $cccccq$ and $bbbbbq$ dibaryon states (in units of MeV).}
\label{table:cccccq+bbbbbq:decay:kci2}
\begin{tabular}{ccccccccccc}
\toprule[1pt]
\toprule[1pt]
System&$J^{P}$&Mass
&$\Omega_{ccc}^{*}\Xi_{cc}^{*}$
&$\Omega_{ccc}^{*}\Xi_{cc}$
&
System&$J^{P}$&Mass
&$\Omega_{bbb}^{*}\Xi_{bb}^{*}$
&$\Omega_{bbb}^{*}\Xi_{bb}$\\
\midrule[1pt]
$cccccn$&$0^{+}$
&$8599$&$140.6$&
&
$bbbbbn$&$0^{+}$
&$24553$&$160.7$\\
&$1^{+}$
&$8516$&$41.8$&$56.3$
&
&$1^{+}$
&$24526$&$63.7$&$66.8$\\
\midrule[1pt]
System&$J^{P}$&Mass
&$\Omega_{ccc}^{*}\Omega_{cc}^{*}$
&$\Omega_{ccc}^{*}\Omega_{cc}$
&
System&$J^{P}$&Mass
&$\Omega_{bbb}^{*}\Omega_{bb}^{*}$
&$\Omega_{bbb}^{*}\Omega_{bb}$\\
\midrule[1pt]
$cccccs$&$0^{+}$
&$8711$&$144.8$&
&
$bbbbbs$&$0^{+}$
&$24624$&$149.2$\\
&$1^{+}$
&$8617$&$38.7$&$57.5$
&
&$1^{+}$
&$24612$&$72.1$&$64.3$\\
\bottomrule[1pt]
\bottomrule[1pt]
\end{tabular}
\end{table*}
%
\begin{table*}
\centering
\caption{The partial width ratios for the $cccccq$ and $bbbbbq$ dibaryon states. For each state, we choose one mode as the reference channel, and the partial width ratios of the other channels are calculated relative to this channel. The masses are all in unit of MeV.}
\label{table:cccccq+bbbbbq:decay:Ratio}
\begin{tabular}{ccccccccccc}
\toprule[1pt]
\toprule[1pt]
System&$J^{P}$&Mass
&$\Omega_{ccc}^{*}\Xi_{cc}^{*}$
&$\Omega_{ccc}^{*}\Xi_{cc}$
&
System&$J^{P}$&Mass
&$\Omega_{bbb}^{*}\Xi_{bb}^{*}$
&$\Omega_{bbb}^{*}\Xi_{bb}$\\
\midrule[1pt]
$cccccn$&$0^{+}$
&$8599$&$1$&
&
$bbbbbn$&$0^{+}$
&$24553$&$1$\\
&$1^{+}$
&$8516$&$1$&$1.3$
&
&$1^{+}$
&$24526$&$1$&$1.0$\\
\midrule[1pt]
System&$J^{P}$&Mass
&$\Omega_{ccc}^{*}\Omega_{cc}^{*}$
&$\Omega_{ccc}^{*}\Omega_{cc}$
&
System&$J^{P}$&Mass
&$\Omega_{bbb}^{*}\Omega_{bb}^{*}$
&$\Omega_{bbb}^{*}\Omega_{bb}$\\
\midrule[1pt]
$cccccs$&$0^{+}$
&$8711$&$1$&
&
$bbbbbs$&$0^{+}$
&$24624$&$1$\\
&$1^{+}$
&$8617$&$1$&$1.5$
&
&$1^{+}$
&$24612$&$1$&$0.9$\\
\bottomrule[1pt]
\bottomrule[1pt]
\end{tabular}
\end{table*}
%


%
%

\begin{table*}[htbp]
\centering
\caption{The eigenvectors of the $ccccbq$ and $bbbbcq$ dibaryon states. The masses are all in units of MeV.}
\label{table:ccccbq+bbbbcq:decay:eigenvector}
\begin{tabular}{cccccccccccccccccccccc}
\toprule[1pt]
\toprule[1pt]
&&&\multicolumn{3}{c}{$ccc{\otimes}cbn$}&\multicolumn{4}{c}{$ccb{\otimes}ccn$}\\
\cmidrule(lr){4-6}
\cmidrule(lr){7-10}
System&$J^{P}$&Mass
&$\Omega_{ccc}^{*}\Xi_{cb}^{*}$
&$\Omega_{ccc}^{*}\Xi_{cb}$
&$\Omega_{ccc}^{*}\Xi_{cb}'$
&$\Omega_{ccb}^{*}\Xi_{cc}^{*}$
&$\Omega_{ccb}^{*}\Xi_{cc}$
&$\Omega_{ccb}\Xi_{cc}^{*}$
&$\Omega_{ccb}\Xi_{cc}$\\
\midrule[1pt]
$ccccbn$&$0^{+}$
&$11693$
&$0.252$&&
&$-0.340$&&&$-0.419$\\
&
&$11840$
&$0.519$&&
&$0.379$&&&$-0.099$\\
&$1^{+}$
&$11698$
&$-0.059$&$-0.068$&$0.233$
&$-0.213$&$0.356$&$-0.304$&$0.166$\\
&
&$11776$
&$-0.270$&$0.289$&$-0.382$
&$-0.307$&$0.128$&$0.113$&$-0.092$\\
&
&$11806$
&$-0.449$&$0.091$&$0.305$
&$0.213$&$0.153$&$0.248$&$0.034$\\
&$2^{+}$
&$11765$
&$0.408$&$-0.377$&$-0.156$
&$-0.272$&$0.136$&$-0.136$\\
\midrule[1pt]
&&&\multicolumn{3}{c}{$ccc{\otimes}cbs$}&\multicolumn{4}{c}{$ccb{\otimes}ccs$}\\
\cmidrule(lr){4-6}
\cmidrule(lr){7-10}
System&$J^{P}$&Mass
&$\Omega_{ccc}^{*}\Omega_{cb}^{*}$
&$\Omega_{ccc}^{*}\Omega_{cb}$
&$\Omega_{ccc}^{*}\Omega_{cb}'$
&$\Omega_{ccb}^{*}\Omega_{cc}^{*}$
&$\Omega_{ccb}^{*}\Omega_{cc}$
&$\Omega_{ccb}\Omega_{cc}^{*}$
&$\Omega_{ccb}\Omega_{cc}$\\
\midrule[1pt]
$ccccbs$&$0^{+}$
&$11793$
&$0.279$&&
&$-0.319$&&&$-0.423$\\
&
&$11941$
&$0.505$&&
&$0.397$&&&$-0.077$\\
&$1^{+}$
&$11802$
&$0.077$&$0.057$&$-0.266$
&$0.190$&$-0.360$&$0.291$&$-0.170$\\
&
&$11871$
&$0.348$&$-0.281$&$0.331$
&$0.268$&$-0.161$&$-0.150$&$0.081$\\
&
&$11907$
&$0.388$&$-0.061$&$-0.350$
&$-0.278$&$-0.104$&$-0.243$&$-0.039$\\
&$2^{+}$
&$11855$
&$0.408$&$-0.385$&$-0.135$
&$-0.272$&$0.136$&$-0.136$\\
\midrule[1pt]
&&&\multicolumn{3}{c}{$bbb{\otimes}cbn$}&\multicolumn{4}{c}{$bbc{\otimes}bbn$}\\
\cmidrule(lr){4-6}
\cmidrule(lr){7-10}
System&$J^{P}$&Mass
&$\Omega_{bbb}^{*}\Xi_{cb}^{*}$
&$\Omega_{bbb}^{*}\Xi_{cb}$
&$\Omega_{bbb}^{*}\Xi_{cb}'$
&$\Omega_{bbc}^{*}\Xi_{bb}^{*}$
&$\Omega_{bbc}^{*}\Xi_{bb}$
&$\Omega_{bbc}\Xi_{bb}^{*}$
&$\Omega_{bbc}\Xi_{bb}$\\
\midrule[1pt]
$bbbbcn$&$0^{+}$
&$21306$
&$-0.567$&&
&$0.098$&&&$-0.331$\\
&
&$21473$
&$0.111$&&
&$-0.500$&&&$-0.275$\\
&$1^{+}$
&$21251$
&$0.022$&$0.571$&$-0.066$
&$0.215$&$0.149$&$-0.130$&$0.167$\\
&
&$21303$
&$-0.526$&$-0.009$&$-0.236$
&$-0.007$&$-0.217$&$-0.254$&$0.011$\\
&
&$21448$
&$0.022$&$-0.052$&$0.017$
&$-0.373$&$0.312$&$-0.292$&$0.096$\\
&$2^{+}$
&$21286$
&$-0.408$&$-0.053$&$-0.405$
&$-0.272$&$0.136$&$-0.136$\\
\midrule[1pt]
&&&\multicolumn{3}{c}{$bbb{\otimes}cbs$}&\multicolumn{4}{c}{$bbc{\otimes}bbs$}\\
\cmidrule(lr){4-6}
\cmidrule(lr){7-10}
System&$J^{P}$&Mass
&$\Omega_{bbb}^{*}\Omega_{cb}^{*}$
&$\Omega_{bbb}^{*}\Omega_{cb}$
&$\Omega_{bbb}^{*}\Omega_{cb}'$
&$\Omega_{bbc}^{*}\Omega_{bb}^{*}$
&$\Omega_{bbc}^{*}\Omega_{bb}$
&$\Omega_{bbc}\Omega_{bb}^{*}$
&$\Omega_{bbc}\Omega_{bb}$\\
\midrule[1pt]
$bbbbcs$&$0^{+}$
&$21398$
&$-0.570$&&
&$0.113$&&&$-0.323$\\
&
&$21550$
&$0.094$&&
&$-0.497$&&&$-0.284$\\
&$1^{+}$
&$21341$
&$0.043$&$0.567$&$-0.088$
&$0.222$&$0.153$&$-0.114$&$0.164$\\
&
&$21395$
&$-0.523$&$-0.001$&$-0.238$
&$-0.016$&$-0.195$&$-0.273$&$0.023$\\
&
&$21524$
&$0.048$&$-0.041$&$0.030$
&$-0.369$&$0.325$&$-0.281$&$0.098$\\
&$2^{+}$
&$21383$
&$-0.408$&$-0.076$&$-0.401$
&$-0.272$&$0.136$&$-0.136$\\
\bottomrule[1pt]
\bottomrule[1pt]
\end{tabular}
\end{table*}
%
\begin{table*}[htbp]
\centering
\caption{The values of $k\cdot|c_{i}|^2$ for the $ccccbq$ and $bbbbcq$ dibaryon states (in units of MeV).}
\label{table:ccccbq+bbbbcq:decay:kci2}
\begin{tabular}{ccccccccccc}
\toprule[1pt]
\toprule[1pt]
&&&\multicolumn{3}{c}{$ccc{\otimes}cbn$}&\multicolumn{4}{c}{$ccb{\otimes}ccn$}\\
\cmidrule(lr){4-6}
\cmidrule(lr){7-10}
System&$J^{P}$&Mass
&$\Omega_{ccc}^{*}\Xi_{cb}^{*}$
&$\Omega_{ccc}^{*}\Xi_{cb}$
&$\Omega_{ccc}^{*}\Xi_{cb}'$
&$\Omega_{ccb}^{*}\Xi_{cc}^{*}$
&$\Omega_{ccb}^{*}\Xi_{cc}$
&$\Omega_{ccb}\Xi_{cc}^{*}$
&$\Omega_{ccb}\Xi_{cc}$\\
\midrule[1pt]
$ccccbn$&$0^{+}$
&$11693$
&$\times$&&
&$\times$&&&$103.8$\\
&
&$11840$
&$183.9$&&
&$113.6$&&&$10.3$\\
&$1^{+}$
&$11698$
&$\times$&$\times$&$\times$
&$\times$&$58.5$&$21.8$&$16.7$\\
&
&$11776$
&$22.6$&$51.7$&$71.2$
&$51.1$&$12.8$&$8.7$&$7.4$\\
&
&$11806$
&$104.2$&$6.2$&$59.8$
&$30.4$&$20.4$&$48.1$&$1.1$\\
&$2^{+}$
&$11765$
&$31.8$&$81.1$&$10.3$
&$36.3$&$13.8$&$11.7$\\
\midrule[1pt]
&&&\multicolumn{3}{c}{$ccc{\otimes}cbs$}&\multicolumn{4}{c}{$ccb{\otimes}ccs$}\\
\cmidrule(lr){4-6}
\cmidrule(lr){7-10}
System&$J^{P}$&Mass
&$\Omega_{ccc}^{*}\Omega_{cb}^{*}$
&$\Omega_{ccc}^{*}\Omega_{cb}$
&$\Omega_{ccc}^{*}\Omega_{cb}'$
&$\Omega_{ccb}^{*}\Omega_{cc}^{*}$
&$\Omega_{ccb}^{*}\Omega_{cc}$
&$\Omega_{ccb}\Omega_{cc}^{*}$
&$\Omega_{ccb}\Omega_{cc}$\\
\midrule[1pt]
$ccccbs$&$0^{+}$
&$11793$
&$\times$&&
&$\times$&&&$108.0$\\
&
&$11941$
&$182.8$&&
&$122.4$&&&$6.3$\\
&$1^{+}$
&$11802$
&$\times$&$0.6$&$\times$
&$\times$&$64.8$&$18.7$&$18.6$\\
&
&$11871$
&$41.1$&$51.6$&$51.7$
&$35.5$&$20.2$&$14.4$&$5.7$\\
&
&$11907$
&$84.9$&$3.0$&$79.8$
&$50.7$&$9.5$&$45.6$&$1.5$\\
&$2^{+}$
&$11855$
&$24.6$&$86.0$&$6.5$
&$29.6$&$13.4$&$10.5$\\
\midrule[1pt]
&&&\multicolumn{3}{c}{$bbb{\otimes}cbn$}&\multicolumn{4}{c}{$bbc{\otimes}bbn$}\\
\cmidrule(lr){4-6}
\cmidrule(lr){7-10}
System&$J^{P}$&Mass
&$\Omega_{bbb}^{*}\Xi_{cb}^{*}$
&$\Omega_{bbb}^{*}\Xi_{cb}$
&$\Omega_{bbb}^{*}\Xi_{cb}'$
&$\Omega_{bbc}^{*}\Xi_{bb}^{*}$
&$\Omega_{bbc}^{*}\Xi_{bb}$
&$\Omega_{bbc}\Xi_{bb}^{*}$
&$\Omega_{bbc}\Xi_{bb}$\\
\midrule[1pt]
$bbbbcn$&$0^{+}$
&$21306$
&$148.7$&&
&$\times$&&&$\times$\\
&
&$21473$
&$16.4$&&
&$241.4$&&&$92.0$\\
&$1^{+}$
&$21251$
&$\times$&$137.5$&$\times$
&$\times$&$\times$&$\times$&$\times$\\
&
&$21303$
&$120.7$&$0.1$&$36.4$
&$\times$&$\times$&$\times$&$\times$\\
&
&$21448$
&$0.6$&$3.9$&$0.4$
&$113.4$&$91.3$&$85.6$&$10.1$\\
&$2^{+}$
&$21286$
&$29.4$&$2.0$&$84.7$
&$\times$&$\times$&$\times$\\
\midrule[1pt]
&&&\multicolumn{3}{c}{$bbb{\otimes}cbs$}&\multicolumn{4}{c}{$bbc{\otimes}bbs$}\\
\cmidrule(lr){4-6}
\cmidrule(lr){7-10}
System&$J^{P}$&Mass
&$\Omega_{bbb}^{*}\Omega_{cb}^{*}$
&$\Omega_{bbb}^{*}\Omega_{cb}$
&$\Omega_{bbb}^{*}\Omega_{cb}'$
&$\Omega_{bbc}^{*}\Omega_{bb}^{*}$
&$\Omega_{bbc}^{*}\Omega_{bb}$
&$\Omega_{bbc}\Omega_{bb}^{*}$
&$\Omega_{bbc}\Omega_{bb}$\\
\midrule[1pt]
$bbbbcs$&$0^{+}$
&$21398$
&$151.5$&&
&$\times$&&&$\times$\\
&
&$21550$
&$11.5$&&
&$236.7$&&&$93.9$\\
&$1^{+}$
&$21341$
&$\times$&$140.5$&$\times$
&$\times$&$\times$&$\times$&$\times$\\
&
&$21395$
&$118.2$&$\sim0$&$34.1$
&$\times$&$\times$&$\times$&$\times$\\
&
&$21524$
&$2.8$&$2.3$&$1.1$
&$109.4$&$90.7$&$78.6$&$9.9$\\
&$2^{+}$
&$21383$
&$43.3$&$4.4$&$79.5$
&$\times$&$\times$&$\times$\\
\bottomrule[1pt]
\bottomrule[1pt]
\end{tabular}
\end{table*}
%
\begin{table*}[htbp]
\centering
\caption{The partial width ratios for the $ccccbq$ and $bbbbcq$ dibaryon states. For each state, we choose one mode as the reference channel, and the partial width ratios of the other channels are calculated relative to this channel. For a dibaryon state decays into two channels through two different quark rearrangements, the $\gamma_{i}$'s, which depend on the wave functions of final states, may not equal (or approximately equal). Thus we compare only the decay width ratios between channels through same quark rearrangement type. The masses are all in unit of MeV.}
\label{table:ccccbq+bbbbcq:decay:Ratio}
\begin{tabular}{cccccccccccccccccccccc}
\toprule[1pt]
\toprule[1pt]
&&&\multicolumn{3}{c}{$ccc{\otimes}cbn$}&\multicolumn{4}{c}{$ccb{\otimes}ccn$}\\
\cmidrule(lr){4-6}
\cmidrule(lr){7-10}
System&$J^{P}$&Mass
&$\Omega_{ccc}^{*}\Xi_{cb}^{*}$
&$\Omega_{ccc}^{*}\Xi_{cb}$
&$\Omega_{ccc}^{*}\Xi_{cb}'$
&$\Omega_{ccb}^{*}\Xi_{cc}^{*}$
&$\Omega_{ccb}^{*}\Xi_{cc}$
&$\Omega_{ccb}\Xi_{cc}^{*}$
&$\Omega_{ccb}\Xi_{cc}$\\
\midrule[1pt]
$ccccbn$&$0^{+}$
&$11693$
&$\times$&&
&$\times$&&&$1$\\
&
&$11840$
&$1$&&
&$11.0$&&&$1$\\
&$1^{+}$
&$11698$
&$\times$&$\times$&$\times$
&$\times$&$3.5$&$1.3$&$1$\\
&
&$11776$
&$1$&$2.3$&$3.1$
&$6.9$&$1.7$&$1.2$&$1$\\
&
&$11806$
&$1$&$0.1$&$0.6$
&$28.0$&$18.8$&$44.4$&$1$\\
&$2^{+}$
&$11765$
&$1$&$2.6$&$0.3$
&$1$&$0.4$&$0.3$\\
\midrule[1pt]
&&&\multicolumn{3}{c}{$ccc{\otimes}cbs$}&\multicolumn{4}{c}{$ccb{\otimes}ccs$}\\
\cmidrule(lr){4-6}
\cmidrule(lr){7-10}
System&$J^{P}$&Mass
&$\Omega_{ccc}^{*}\Omega_{cb}^{*}$
&$\Omega_{ccc}^{*}\Omega_{cb}$
&$\Omega_{ccc}^{*}\Omega_{cb}'$
&$\Omega_{ccb}^{*}\Omega_{cc}^{*}$
&$\Omega_{ccb}^{*}\Omega_{cc}$
&$\Omega_{ccb}\Omega_{cc}^{*}$
&$\Omega_{ccb}\Omega_{cc}$\\
\midrule[1pt]
$ccccbs$&$0^{+}$
&$11793$
&$\times$&&
&$\times$&&&$1$\\
&
&$11941$
&$1$&&
&$19.5$&&&$1$\\
&$1^{+}$
&$11802$
&$\times$&$1$&$\times$
&$\times$&$3.5$&$1.0$&$1$\\
&
&$11871$
&$0.8$&$1$&$1.0$
&$6.2$&$3.5$&$2.5$&$1$\\
&
&$11907$
&$28.3$&$1$&$26.6$
&$34.9$&$6.6$&$31.4$&$1$\\
&$2^{+}$
&$11855$
&$1$&$3.5$&$0.3$
&$1$&$0.5$&$0.4$\\
\midrule[1pt]
&&&\multicolumn{3}{c}{$bbb{\otimes}cbn$}&\multicolumn{4}{c}{$bbc{\otimes}bbn$}\\
\cmidrule(lr){4-6}
\cmidrule(lr){7-10}
System&$J^{P}$&Mass
&$\Omega_{bbb}^{*}\Xi_{cb}^{*}$
&$\Omega_{bbb}^{*}\Xi_{cb}$
&$\Omega_{bbb}^{*}\Xi_{cb}'$
&$\Omega_{bbc}^{*}\Xi_{bb}^{*}$
&$\Omega_{bbc}^{*}\Xi_{bb}$
&$\Omega_{bbc}\Xi_{bb}^{*}$
&$\Omega_{bbc}\Xi_{bb}$\\
\midrule[1pt]
$bbbbcn$&$0^{+}$
&$21306$
&$1$&&
&$\times$&&&$\times$\\
&
&$21473$
&$1$&&
&$2.6$&&&$1$\\
&$1^{+}$
&$21251$
&$\times$&$1$&$\times$
&$\times$&$\times$&$\times$&$\times$\\
&
&$21303$
&$3.3$&$\sim0$&$1$
&$\times$&$\times$&$\times$&$\times$\\
&
&$21448$
&$0.2$&$1$&$0.1$
&$11.2$&$9.0$&$8.5$&$1$\\
&$2^{+}$
&$21286$
&$14.4$&$1$&$41.6$
&$\times$&$\times$&$\times$\\
\midrule[1pt]
&&&\multicolumn{3}{c}{$bbb{\otimes}cbs$}&\multicolumn{4}{c}{$bbc{\otimes}bbs$}\\
\cmidrule(lr){4-6}
\cmidrule(lr){7-10}
System&$J^{P}$&Mass
&$\Omega_{bbb}^{*}\Omega_{cb}^{*}$
&$\Omega_{bbb}^{*}\Omega_{cb}$
&$\Omega_{bbb}^{*}\Omega_{cb}'$
&$\Omega_{bbc}^{*}\Omega_{bb}^{*}$
&$\Omega_{bbc}^{*}\Omega_{bb}$
&$\Omega_{bbc}\Omega_{bb}^{*}$
&$\Omega_{bbc}\Omega_{bb}$\\
\midrule[1pt]
$bbbbcs$&$0^{+}$
&$21398$
&$1$&&
&$\times$&&&$\times$\\
&
&$21550$
&$1$&&
&$2.5$&&&$1$\\
&$1^{+}$
&$21341$
&$\times$&$1$&$\times$
&$\times$&$\times$&$\times$&$\times$\\
&
&$21395$
&$3.5$&$\sim0$&$1$
&$\times$&$\times$&$\times$&$\times$\\
&
&$21524$
&$2.5$&$2.1$&$1$
&$11.0$&$9.1$&$7.9$&$1$\\
&$2^{+}$
&$21383$
&$9.8$&$1$&$18.0$
&$\times$&$\times$&$\times$\\
\bottomrule[1pt]
\bottomrule[1pt]
\end{tabular}
\end{table*}
%


%
%

\begin{table*}[htbp]
\centering
\caption{The eigenvectors of the $cccbbq$ dibaryon states. The scattering states are marked with a dagger ($\dagger$). The masses are all in units of MeV.}
\label{table:cccbbq:decay:eigenvector}
\begin{tabular}{ccccccccccccccccccc}
\toprule[1pt]
\toprule[1pt]
&&&\multicolumn{2}{c}{$ccc{\otimes}bbn$}&\multicolumn{6}{c}{$ccb{\otimes}cbn$}&\multicolumn{4}{c}{$ccn{\otimes}bbc$}\\
\cmidrule(lr){4-5}
\cmidrule(lr){6-11}
\cmidrule(lr){12-15}
System&$J^{P}$&Mass
&$\Omega_{ccc}^{*}\Xi_{bb}^{*}$
&$\Omega_{ccc}^{*}\Xi_{bb}$
&$\Omega_{ccb}^{*}\Xi_{cb}^{*}$
&$\Omega_{ccb}^{*}\Xi_{cb}$
&$\Omega_{ccb}^{*}\Xi_{cb}'$
&$\Omega_{ccb}\Xi_{cb}^{*}$
&$\Omega_{ccb}\Xi_{cb}$
&$\Omega_{ccb}\Xi_{cb}'$
&$\Xi_{cc}^{*}\Omega_{bbc}^{*}$
&$\Xi_{cc}^{*}\Omega_{bbc}$
&$\Xi_{cc}\Omega_{bbc}^{*}$
&$\Xi_{cc}\Omega_{bbc}$\\
\midrule[1pt]
$cccbbn$&$0^{+}$
&$14887$
&$0.369$&
&$0.064$&&&&$0.253$&$-0.212$
&$0.541$&&&$-0.322$\\
&
&$14981^{\dagger}$
&$0.899$&
&$-0.017$&&&&$0.220$&$0.284$
&$-0.308$&&&$-0.209$\\
&
&$15093$
&$0.237$&
&$-0.505$&&&&$-0.003$&$-0.242$
&$-0.142$&&&$-0.034$\\
&$1^{+}$
&$14833$
&$0.092$&$0.073$
&$0.177$&$0.029$&$-0.209$&$0.166$&$0.029$&$-0.090$
&$0.086$&$0.049$&$-0.487$&$-0.437$\\
&
&$14884$
&$-0.266$&$-0.241$
&$-0.000$&$-0.120$&$0.083$&$0.190$&$-0.234$&$0.030$
&$-0.471$&$-0.251$&$-0.288$&$0.174$\\
&
&$14953^{\dagger}$
&$0.357$&$-0.890$
&$0.010$&$-0.075$&$-0.080$&$0.035$&$0.102$&$0.320$
&$0.199$&$-0.261$&$-0.009$&$-0.023$\\
&
&$14984$
&$-0.859$&$-0.336$
&$-0.036$&$-0.096$&$-0.206$&$-0.173$&$-0.178$&$-0.021$
&$0.295$&$0.148$&$-0.138$&$0.159$\\
&
&$15039$
&$-0.236$&$0.177$
&$0.374$&$-0.265$&$0.187$&$-0.170$&$0.159$&$0.112$
&$-0.051$&$0.071$&$-0.078$&$0.014$\\
&$2^{+}$
&$14872$
&$-0.097$&$-0.303$
&$0.166$&$-0.264$&$-0.015$&$0.117$&&
&$-0.365$&$-0.433$&$-0.300$\\
&
&$14962^{\dagger}$
&$0.234$&$-0.933$
&$-0.139$&$-0.047$&$0.272$&$0.125$&&
&$0.355$&$-0.034$&$-0.095$\\
&
&$14976^{\dagger}$
&$0.967$&$0.195$
&$0.165$&$0.197$&$0.021$&$0.211$&&
&$-0.008$&$-0.265$&$0.222$\\
&$3^{+}$
&$14974^{\dagger}$
&$1$&
&$0.333$&&&&&
&$0.333$\\
\midrule[1pt]
&&&\multicolumn{2}{c}{$ccc{\otimes}bbs$}&\multicolumn{6}{c}{$ccb{\otimes}cbs$}&\multicolumn{4}{c}{$ccs{\otimes}bbc$}\\
\cmidrule(lr){4-5}
\cmidrule(lr){6-11}
\cmidrule(lr){12-15}
System&$J^{P}$&Mass
&$\Omega_{ccc}^{*}\Omega_{bb}^{*}$
&$\Omega_{ccc}^{*}\Omega_{bb}$
&$\Omega_{ccb}^{*}\Omega_{cb}^{*}$
&$\Omega_{ccb}^{*}\Omega_{cb}$
&$\Omega_{ccb}^{*}\Omega_{cb}'$
&$\Omega_{ccb}\Omega_{cb}^{*}$
&$\Omega_{ccb}\Omega_{cb}$
&$\Omega_{ccb}\Omega_{cb}'$
&$\Omega_{cc}^{*}\Omega_{bbc}^{*}$
&$\Omega_{cc}^{*}\Omega_{bbc}$
&$\Omega_{cc}\Omega_{bbc}^{*}$
&$\Omega_{cc}\Omega_{bbc}$\\
\midrule[1pt]
$cccbbs$&$0^{+}$
&$14986$
&$0.507$&
&$0.058$&&&&$0.275$&$-0.181$
&$0.485$&&&$-0.351$\\
&
&$15069$
&$0.835$&
&$-0.040$&&&&$0.194$&$0.297$
&$-0.393$&&&$-0.156$\\
&
&$15187$
&$0.214$&
&$-0.504$&&&&$-0.022$&$-0.248$
&$-0.133$&&&$-0.028$\\
&$1^{+}$
&$14934$
&$-0.120$&$-0.084$
&$-0.181$&$-0.020$&$0.206$&$-0.164$&$-0.036$&$0.090$
&$-0.092$&$-0.051$&$0.477$&$0.445$\\
&
&$14984$
&$-0.371$&$-0.271$
&$-0.005$&$-0.123$&$0.055$&$0.173$&$-0.255$&$0.033$
&$-0.428$&$-0.227$&$-0.316$&$0.181$\\
&
&$15041$
&$0.465$&$-0.844$
&$0.010$&$-0.065$&$-0.045$&$0.065$&$0.134$&$0.313$
&$0.142$&$-0.287$&$0.007$&$-0.035$\\
&
&$15072$
&$-0.773$&$-0.411$
&$-0.004$&$-0.120$&$-0.208$&$-0.197$&$-0.117$&$0.034$
&$0.373$&$0.145$&$-0.116$&$0.123$\\
&
&$15127$
&$0.182$&$-0.195$
&$-0.374$&$0.248$&$-0.220$&$0.154$&$-0.171$&$-0.105$
&$0.083$&$-0.055$&$0.070$&$-0.011$\\
&$2^{+}$
&$14976$
&$-0.154$&$-0.382$
&$0.143$&$-0.280$&$0.019$&$0.109$&&
&$-0.337$&$-0.413$&$-0.322$\\
&
&$15051$
&$0.764$&$-0.631$
&$-0.015$&$0.095$&$0.226$&$0.225$&&
&$0.285$&$-0.180$&$0.060$\\
&
&$15061$
&$-0.626$&$-0.676$
&$-0.231$&$-0.149$&$0.154$&$-0.107$&&
&$0.254$&$0.237$&$-0.203$\\
&$3^{+}$
&$15053^{\dagger}$
&$1$&
&$0.333$&&&&&
&$0.333$\\
\bottomrule[1pt]
\bottomrule[1pt]
\end{tabular}
\end{table*}
%
\begin{table*}[htbp]
\centering
\caption{The values of $k\cdot|c_{i}|^2$ for the $cccbbq$ dibaryon states (in units of MeV). The scattering states are marked with a dagger ($\dagger$).}
\label{table:cccbbq:decay:kci2}
\begin{tabular}{cccccccccccccccccccc}
\toprule[1pt]
\toprule[1pt]
&&&\multicolumn{2}{c}{$ccc{\otimes}bbn$}&\multicolumn{6}{c}{$ccb{\otimes}cbn$}&\multicolumn{4}{c}{$ccn{\otimes}bbc$}\\
\cmidrule(lr){4-5}
\cmidrule(lr){6-11}
\cmidrule(lr){12-15}
System&$J^{P}$&Mass
&$\Omega_{ccc}^{*}\Xi_{bb}^{*}$
&$\Omega_{ccc}^{*}\Xi_{bb}$
&$\Omega_{ccb}^{*}\Xi_{cb}^{*}$
&$\Omega_{ccb}^{*}\Xi_{cb}$
&$\Omega_{ccb}^{*}\Xi_{cb}'$
&$\Omega_{ccb}\Xi_{cb}^{*}$
&$\Omega_{ccb}\Xi_{cb}$
&$\Omega_{ccb}\Xi_{cb}'$
&$\Xi_{cc}^{*}\Omega_{bbc}^{*}$
&$\Xi_{cc}^{*}\Omega_{bbc}$
&$\Xi_{cc}\Omega_{bbc}^{*}$
&$\Xi_{cc}\Omega_{bbc}$\\
\midrule[1pt]
$cccbbn$&$0^{+}$
&$14887$
&$\times$&
&$\times$&&&&$\times$&$\times$
&$\times$&&&$72.3$\\
&
&$14981^{\dagger}$
&$167.0$&
&$\times$&&&&$34.5$&$45.5$
&$66.8$&&&$44.0$\\
&
&$15093$
&$49.3$&
&$218.2$&&&&$0.01$&$62.8$
&$21.3$&&&$1.4$\\
&$1^{+}$
&$14833$
&$\times$&$\times$
&$\times$&$\times$&$\times$&$\times$&$\times$&$\times$
&$\times$&$\times$&$32.4$&$83.5$\\
&
&$14884$
&$\times$&$\times$
&$\times$&$\times$&$\times$&$\times$&$\times$&$\times$
&$\times$&$22.5$&$45.4$&$20.8$\\
&
&$14953^{\dagger}$
&$\times$&$\times$
&$\times$&$1.4$&$\times$&$\times$&$5.7$&$33.9$
&$23.1$&$48.8$&$0.1$&$0.5$\\
&
&$14984$
&$180.1$&$49.1$
&$\times$&$5.0$&$13.7$&$11.5$&$23.0$&$0.3$
&$62.0$&$18.1$&$17.6$&$25.6$\\
&
&$15039$
&$36.0$&$23.2$
&$79.9$&$58.8$&$25.2$&$21.5$&$24.4$&$10.9$
&$2.3$&$5.1$&$6.6$&$0.2$\\
&$2^{+}$
&$14872$
&$\times$&$\times$
&$\times$&$\times$&$\times$&$\times$&&
&$\times$&$47.2$&$43.5$\\
&
&$14962^{\dagger}$
&$\times$&$194.6$
&$\times$&$0.8$&$\times$&$\times$&&
&$78.6$&$0.9$&$7.7$\\
&
&$14976^{\dagger}$
&$99.8$&$14.3$
&$\times$&$19.0$&$0.1$&$13.7$&&
&$0.04$&$56.4$&$44.6$\\
&$3^{+}$
&$14974^{\dagger}$
&$0.05$&
&$\times$&&&&&
&$75.2$\\
\midrule[1pt]
&&&\multicolumn{2}{c}{$ccc{\otimes}bbs$}&\multicolumn{6}{c}{$ccb{\otimes}cbs$}&\multicolumn{4}{c}{$ccs{\otimes}bbc$}\\
\cmidrule(lr){4-5}
\cmidrule(lr){6-11}
\cmidrule(lr){12-15}
System&$J^{P}$&Mass
&$\Omega_{ccc}^{*}\Omega_{bb}^{*}$
&$\Omega_{ccc}^{*}\Omega_{bb}$
&$\Omega_{ccb}^{*}\Omega_{cb}^{*}$
&$\Omega_{ccb}^{*}\Omega_{cb}$
&$\Omega_{ccb}^{*}\Omega_{cb}'$
&$\Omega_{ccb}\Omega_{cb}^{*}$
&$\Omega_{ccb}\Omega_{cb}$
&$\Omega_{ccb}\Omega_{cb}'$
&$\Omega_{cc}^{*}\Omega_{bbc}^{*}$
&$\Omega_{cc}^{*}\Omega_{bbc}$
&$\Omega_{cc}\Omega_{bbc}^{*}$
&$\Omega_{cc}\Omega_{bbc}$\\
\midrule[1pt]
$cccbbs$&$0^{+}$
&$14986$
&$\times$&
&$\times$&&&&$\times$&$\times$
&$\times$&&&$87.1$\\
&
&$15069$
&$225.8$&
&$\times$&&&&$26.9$&$43.2$
&$97.9$&&&$24.1$\\
&
&$15187$
&$42.7$&
&$219.8$&&&&$0.6$&$65.3$
&$18.5$&&&$1.0$\\
&$1^{+}$
&$14934$
&$\times$&$\times$
&$\times$&$\times$&$\times$&$\times$&$\times$&$\times$
&$\times$&$\times$&$42.3$&$90.8$\\
&
&$14984$
&$\times$&$\times$
&$\times$&$\times$&$\times$&$\times$&$\times$&$\times$
&$\times$&$15.8$&$55.9$&$22.9$\\
&
&$15041$
&$\times$&$\times$
&$\times$&$1.1$&$\times$&$\times$&$9.9$&$16.5$
&$9.9$&$53.5$&$0.03$&$1.1$\\
&
&$15072$
&$210.9$&$71.5$
&$\times$&$7.9$&$6.9$&$13.5$&$10.0$&$0.6$
&$89.9$&$16.3$&$12.2$&$15.2$\\
&
&$15127$
&$23.1$&$28.1$
&$76.6$&$52.0$&$32.0$&$17.4$&$28.5$&$9.0$
&$6.0$&$2.9$&$5.2$&$0.1$\\
&$2^{+}$
&$14976$
&$\times$&$\times$
&$\times$&$\times$&$\times$&$\times$&&
&$\times$&$37.8$&$53.6$\\
&
&$15051$
&$\times$&$79.7$
&$\times$&$3.4$&$\times$&$\times$&&
&$44.2$&$22.3$&$3.0$\\
&
&$15061$
&$90.3$&$150.3$
&$\times$&$10.4$&$\times$&$2.3$&&
&$38.4$&$41.2$&$35.6$\\
&$3^{+}$
&$15053^{\dagger}$
&$\times$&
&$\times$&&&&&
&$61.8$\\
\bottomrule[1pt]
\bottomrule[1pt]
\end{tabular}
\end{table*}
%
\begin{table*}[htbp]
\centering
\caption{The partial width ratios for the $cccbbq$ dibaryon states. For each state, we choose one mode as the reference channel, and the partial width ratios of the other channels are calculated relative to this channel. For a dibaryon state decays into two channels through two different quark rearrangements, the $\gamma_{i}$'s, which depend on the wave functions of final states, may not equal (or approximately equal). Thus we compare only the decay width ratios between channels through same quark rearrangement type. The scattering states are marked with a dagger ($\dagger$). The masses are all in unit of MeV.}
\label{table:cccbbq:decay:Ratio}
\begin{tabular}{cccccccccccccccccccccc}
\toprule[1pt]
\toprule[1pt]
&&&\multicolumn{2}{c}{$ccc{\otimes}bbn$}&\multicolumn{6}{c}{$ccb{\otimes}cbn$}&\multicolumn{4}{c}{$ccn{\otimes}bbc$}\\
\cmidrule(lr){4-5}
\cmidrule(lr){6-11}
\cmidrule(lr){12-15}
System&$J^{P}$&Mass
&$\Omega_{ccc}^{*}\Xi_{bb}^{*}$
&$\Omega_{ccc}^{*}\Xi_{bb}$
&$\Omega_{ccb}^{*}\Xi_{cb}^{*}$
&$\Omega_{ccb}^{*}\Xi_{cb}$
&$\Omega_{ccb}^{*}\Xi_{cb}'$
&$\Omega_{ccb}\Xi_{cb}^{*}$
&$\Omega_{ccb}\Xi_{cb}$
&$\Omega_{ccb}\Xi_{cb}'$
&$\Xi_{cc}^{*}\Omega_{bbc}^{*}$
&$\Xi_{cc}^{*}\Omega_{bbc}$
&$\Xi_{cc}\Omega_{bbc}^{*}$
&$\Xi_{cc}\Omega_{bbc}$\\
\midrule[1pt]
$cccbbn$&$0^{+}$
&$14887$
&$\times$&
&$\times$&&&&$\times$&$\times$
&$\times$&&&$1$\\
&
&$14981^{\dagger}$
&$1$&
&$\times$&&&&$0.8$&$1$
&$1.5$&&&$1$\\
&
&$15093$
&$1$&
&$3.5$&&&&$\sim0$&$1$
&$14.7$&&&$1$\\
&$1^{+}$
&$14833$
&$\times$&$\times$
&$\times$&$\times$&$\times$&$\times$&$\times$&$\times$
&$\times$&$\times$&$0.4$&$1$\\
&
&$14884$
&$\times$&$\times$
&$\times$&$\times$&$\times$&$\times$&$\times$&$\times$
&$\times$&$1.1$&$2.2$&$1$\\
&
&$14953^{\dagger}$
&$\times$&$\times$
&$\times$&$1$&$\times$&$\times$&$3.9$&$23.6$
&$1$&$2.1$&$\sim0$&$\sim0$\\
&
&$14984$
&$1$&$0.3$
&$\times$&$1$&$2.7$&$2.3$&$4.6$&$0.1$
&$2.4$&$0.7$&$0.7$&$1$\\
&
&$15039$
&$1$&$0.6$
&$7.4$&$5.4$&$2.3$&$2.0$&$2.2$&$1$
&$1$&$2.2$&$2.8$&$0.1$\\
&$2^{+}$
&$14872$
&$\times$&$\times$
&$\times$&$\times$&$\times$&$\times$&&
&$\times$&$1.1$&$1$\\
&
&$14962^{\dagger}$
&$\times$&$1$
&$\times$&$1$&$\times$&$\times$&&
&$10.3$&$0.1$&$1$\\
&
&$14976^{\dagger}$
&$7.0$&$1$
&$\times$&$1$&$\sim0$&$0.7$&&
&$\sim0$&$1.3$&$1$\\
&$3^{+}$
&$14974^{\dagger}$
&$1$&
&$\times$&&&&&
&$1$\\
\midrule[1pt]
&&&\multicolumn{2}{c}{$ccc{\otimes}bbs$}&\multicolumn{6}{c}{$ccb{\otimes}cbs$}&\multicolumn{4}{c}{$ccs{\otimes}bbc$}\\
\cmidrule(lr){4-5}
\cmidrule(lr){6-11}
\cmidrule(lr){12-15}
System&$J^{P}$&Mass
&$\Omega_{ccc}^{*}\Omega_{bb}^{*}$
&$\Omega_{ccc}^{*}\Omega_{bb}$
&$\Omega_{ccb}^{*}\Omega_{cb}^{*}$
&$\Omega_{ccb}^{*}\Omega_{cb}$
&$\Omega_{ccb}^{*}\Omega_{cb}'$
&$\Omega_{ccb}\Omega_{cb}^{*}$
&$\Omega_{ccb}\Omega_{cb}$
&$\Omega_{ccb}\Omega_{cb}'$
&$\Omega_{cc}^{*}\Omega_{bbc}^{*}$
&$\Omega_{cc}^{*}\Omega_{bbc}$
&$\Omega_{cc}\Omega_{bbc}^{*}$
&$\Omega_{cc}\Omega_{bbc}$\\
\midrule[1pt]
$cccbbs$&$0^{+}$
&$14986$
&$\times$&
&$\times$&&&&$\times$&$\times$
&$\times$&&&$1$\\
&
&$15069$
&$1$&
&$\times$&&&&$0.6$&$1$
&$4.1$&&&$1$\\
&
&$15187$
&$1$&
&$3.4$&&&&$0.01$&$1$
&$18.0$&&&$1$\\
&$1^{+}$
&$14934$
&$\times$&$\times$
&$\times$&$\times$&$\times$&$\times$&$\times$&$\times$
&$\times$&$\times$&$0.5$&$1$\\
&
&$14984$
&$\times$&$\times$
&$\times$&$\times$&$\times$&$\times$&$\times$&$\times$
&$\times$&$0.7$&$2.4$&$1$\\
&
&$15041$
&$\times$&$\times$
&$\times$&$1$&$\times$&$\times$&$9.3$&$15.5$
&$8.7$&$47.0$&$0.03$&$1$\\
&
&$15072$
&$2.9$&$1$
&$\times$&$1$&$0.9$&$1.7$&$1.3$&$0.1$
&$5.9$&$1.1$&$0.8$&$1$\\
&
&$15127$
&$0.8$&$1$
&$8.5$&$5.7$&$3.5$&$1.9$&$3.2$&$1$
&$2.1$&$1$&$1.8$&$0.05$\\
&$2^{+}$
&$14976$
&$\times$&$\times$
&$\times$&$\times$&$\times$&$\times$&&
&$\times$&$0.7$&$1$\\
&
&$15051$
&$\times$&$1$
&$\times$&$1$&$\times$&$\times$&&
&$14.9$&$7.5$&$1$\\
&
&$15061$
&$0.6$&$1$
&$\times$&$1$&$\times$&$0.2$&&
&$1.1$&$1.2$&$1$\\
&$3^{+}$
&$15053^{\dagger}$
&$\times$&
&$\times$&&&&&
&$1$\\
\bottomrule[1pt]
\bottomrule[1pt]
\end{tabular}
\end{table*}
%
\begin{table*}[htbp]
\centering
\caption{The eigenvectors of the $bbbccq$ dibaryon states. The scattering states are marked with a dagger ($\dagger$). The masses are all in units of MeV.}
\label{table:bbbccq:decay:eigenvector}
\begin{tabular}{cccccccccccccccccccccc}
\toprule[1pt]
\toprule[1pt]
&&&\multicolumn{2}{c}{$bbb{\otimes}ccn$}&\multicolumn{6}{c}{$bbc{\otimes}cbn$}&\multicolumn{4}{c}{$bbn{\otimes}ccb$}\\
\cmidrule(lr){4-5}
\cmidrule(lr){6-11}
\cmidrule(lr){12-15}
System&$J^{P}$&Mass
&$\Omega_{bbb}^{*}\Xi_{cc}^{*}$
&$\Omega_{bbb}^{*}\Xi_{cc}$
&$\Omega_{bbc}^{*}\Xi_{cb}^{*}$
&$\Omega_{bbc}^{*}\Xi_{cb}$
&$\Omega_{bbc}^{*}\Xi_{cb}'$
&$\Omega_{bbc}\Xi_{cb}^{*}$
&$\Omega_{bbc}\Xi_{cb}$
&$\Omega_{bbc}\Xi_{cb}'$
&$\Xi_{bb}^{*}\Omega_{ccb}^{*}$
&$\Xi_{bb}^{*}\Omega_{ccb}$
&$\Xi_{bb}\Omega_{ccb}^{*}$
&$\Xi_{bb}\Omega_{ccb}$\\
\midrule[1pt]
$bbbccn$&$0^{+}$
&$18003^{\dagger}$
&$0.993$&
&$0.051$&&&&$0.061$&$0.328$
&$-0.081$&&&$-0.322$\\
&
&$18167$
&$-0.077$&
&$-0.361$&&&&$0.395$&$0.049$
&$-0.269$&&&$0.122$\\
&
&$18280$
&$0.090$&
&$0.356$&&&&$0.120$&$-0.111$
&$-0.573$&&&$0.171$\\
&$1^{+}$
&$17941^{\dagger}$
&$-0.009$&$0.996$
&$-0.063$&$0.082$&$0.127$&$0.032$&$-0.289$&$-0.030$
&$-0.140$&$0.271$&$0.066$&$-0.119$\\
&
&$18004^{\dagger}$
&$-0.994$&$-0.013$
&$0.004$&$-0.033$&$-0.146$&$0.149$&$-0.035$&$-0.259$
&$0.019$&$0.152$&$-0.152$&$0.254$\\
&
&$18161$
&$0.071$&$-0.073$
&$0.252$&$-0.311$&$0.153$&$-0.218$&$-0.124$&$-0.224$
&$0.041$&$-0.017$&$-0.196$&$-0.167$\\
&
&$18233$
&$0.074$&$0.003$
&$0.320$&$0.101$&$0.142$&$0.052$&$0.016$&$-0.091$
&$-0.118$&$-0.063$&$0.476$&$0.365$\\
&
&$18250$
&$-0.024$&$-0.040$
&$0.054$&$-0.192$&$0.040$&$0.224$&$-0.040$&$0.135$
&$0.568$&$0.244$&$0.232$&$-0.087$\\
&$2^{+}$
&$17942^{\dagger}$
&$0.003$&$-0.999$
&$0.102$&$0.278$&$-0.101$&$-0.116$&&
&$0.204$&$-0.130$&$-0.231$\\
&
&$18006^{\dagger}$
&$1.000$&$0.003$
&$-0.115$&$0.028$&$0.217$&$-0.224$&&
&$0.103$&$-0.229$&$0.219$\\
&
&$18230$
&$-0.017$&$0.041$
&$-0.225$&$-0.041$&$-0.220$&$-0.103$&&
&$-0.455$&$-0.436$&$-0.217$\\
&$3^{+}$
&$18006^{\dagger}$
&$1$&
&$-0.333$&&&&&
&$0.333$\\
\midrule[1pt]
&&&\multicolumn{2}{c}{$bbb{\otimes}ccs$}&\multicolumn{6}{c}{$bbc{\otimes}cbs$}&\multicolumn{4}{c}{$bbs{\otimes}ccb$}\\
\cmidrule(lr){4-5}
\cmidrule(lr){6-11}
\cmidrule(lr){12-15}
System&$J^{P}$&Mass
&$\Omega_{bbb}^{*}\Omega_{cc}^{*}$
&$\Omega_{bbb}^{*}\Omega_{cc}$
&$\Omega_{bbc}^{*}\Omega_{cb}^{*}$
&$\Omega_{bbc}^{*}\Omega_{cb}$
&$\Omega_{bbc}^{*}\Omega_{cb}'$
&$\Omega_{bbc}\Omega_{cb}^{*}$
&$\Omega_{bbc}\Omega_{cb}$
&$\Omega_{bbc}\Omega_{cb}'$
&$\Omega_{bb}^{*}\Omega_{ccb}^{*}$
&$\Omega_{bb}^{*}\Omega_{ccb}$
&$\Omega_{bb}\Omega_{ccb}^{*}$
&$\Omega_{bb}\Omega_{ccb}$\\
\midrule[1pt]
$bbbccs$&$0^{+}$
&$18109^{\dagger}$
&$0.992$&
&$0.047$&&&&$0.091$&$0.323$
&$-0.095$&&&$-0.317$\\
&
&$18251$
&$-0.098$&
&$-0.343$&&&&$0.401$&$0.012$
&$-0.296$&&&$0.140$\\
&
&$18362$
&$0.083$&
&$0.374$&&&&$0.091$&$-0.122$
&$-0.557$&&&$0.168$\\
&$1^{+}$
&$18040^{\dagger}$
&$-0.011$&$0.997$
&$-0.063$&$0.089$&$0.122$&$0.031$&$-0.290$&$-0.016$
&$-0.142$&$0.270$&$0.064$&$-0.118$\\
&
&$18110^{\dagger}$
&$-0.994$&$-0.016$
&$0.006$&$-0.046$&$-0.142$&$0.152$&$-0.050$&$-0.255$
&$0.030$&$0.156$&$-0.149$&$0.251$\\
&
&$18256$
&$0.069$&$-0.075$
&$0.226$&$-0.315$&$0.160$&$-0.216$&$-0.137$&$-0.204$
&$0.065$&$-0.006$&$-0.229$&$-0.200$\\
&
&$18323$
&$0.078$&$0.004$
&$0.322$&$0.119$&$0.136$&$-0.015$&$0.006$&$-0.138$
&$-0.231$&$-0.114$&$0.397$&$0.360$\\
&
&$18328$
&$0.010$&$-0.033$
&$0.119$&$-0.159$&$0.080$&$0.228$&$-0.029$&$0.120$
&$0.529$&$0.224$&$0.331$&$-0.012$\\
&$2^{+}$
&$18041^{\dagger}$
&$0.003$&$-0.9995$
&$0.104$&$0.272$&$-0.114$&$-0.115$&&
&$0.209$&$-0.125$&$-0.228$\\
&
&$18112^{\dagger}$
&$0.999$&$0.004$
&$-0.120$&$0.038$&$0.210$&$-0.226$&&
&$0.093$&$-0.239$&$0.214$\\
&
&$18308$
&$-0.039$&$0.030$
&$-0.221$&$-0.051$&$-0.223$&$-0.099$&&
&$-0.455$&$-0.432$&$-0.224$\\
&$3^{+}$
&$18112^{\dagger}$
&$1$&
&$-0.333$&&&&&
&$0.333$\\
\bottomrule[1pt]
\bottomrule[1pt]
\end{tabular}
\end{table*}
%
\begin{table*}[htbp]
\centering
\caption{The values of $k\cdot|c_{i}|^2$ for the $bbbccq$ dibaryon states (in units of MeV). The scattering states are marked with a dagger ($\dagger$).}
\label{table:bbbccq:decay:kci2}
\begin{tabular}{cccccccccccccccccccccc}
\toprule[1pt]
\toprule[1pt]
&&&\multicolumn{2}{c}{$bbb{\otimes}ccn$}&\multicolumn{6}{c}{$bbc{\otimes}cbn$}&\multicolumn{4}{c}{$bbn{\otimes}ccb$}\\
\cmidrule(lr){4-5}
\cmidrule(lr){6-11}
\cmidrule(lr){12-15}
System&$J^{P}$&Mass
&$\Omega_{bbb}^{*}\Xi_{cc}^{*}$
&$\Omega_{bbb}^{*}\Xi_{cc}$
&$\Omega_{bbc}^{*}\Xi_{cb}^{*}$
&$\Omega_{bbc}^{*}\Xi_{cb}$
&$\Omega_{bbc}^{*}\Xi_{cb}'$
&$\Omega_{bbc}\Xi_{cb}^{*}$
&$\Omega_{bbc}\Xi_{cb}$
&$\Omega_{bbc}\Xi_{cb}'$
&$\Xi_{bb}^{*}\Omega_{ccb}^{*}$
&$\Xi_{bb}^{*}\Omega_{ccb}$
&$\Xi_{bb}\Omega_{ccb}^{*}$
&$\Xi_{bb}\Omega_{ccb}$\\
\midrule[1pt]
$bbbccn$&$0^{+}$
&$18003^{\dagger}$
&$\times$&
&$\times$&&&&$\times$&$\times$
&$\times$&&&$\times$\\
&
&$18167$
&$5.9$&
&$\times$&&&&$129.0$&$1.6$
&$\times$&&&$3.9$\\
&
&$18280$
&$10.4$&
&$123.4$&&&&$18.5$&$14.6$
&$259.7$&&&$30.4$\\
&$1^{+}$
&$17941^{\dagger}$
&$\times$&$\times$
&$\times$&$\times$&$\times$&$\times$&$\times$&$\times$
&$\times$&$\times$&$\times$&$\times$\\
&
&$18004^{\dagger}$
&$\times$&$0.1$
&$\times$&$\times$&$\times$&$\times$&$\times$&$\times$
&$\times$&$\times$&$\times$&$\times$\\
&
&$18161$
&$4.8$&$6.0$
&$\times$&$58.6$&$9.0$&$21.2$&$12.3$&$32.5$
&$\times$&$\times$&$\times$&$3.9$\\
&
&$18233$
&$6.3$&$0.01$
&$75.3$&$10.0$&$17.7$&$2.4$&$0.3$&$8.4$
&$6.2$&$2.7$&$138.8$&$107.8$\\
&
&$18250$
&$0.7$&$2.1$
&$2.5$&$39.3$&$1.5$&$49.2$&$1.9$&$19.7$
&$191.9$&$47.6$&$39.1$&$6.9$\\
&$2^{+}$
&$17942^{\dagger}$
&$\times$&$\times$
&$\times$&$\times$&$\times$&$\times$&&
&$\times$&$\times$&$\times$\\
&
&$18006^{\dagger}$
&$\times$&$0.01$
&$\times$&$\times$&$\times$&$\times$&&
&$\times$&$\times$&$\times$\\
&
&$18230$
&$0.3$&$2.2$
&$36.5$&$1.7$&$41.6$&$9.4$&&
&$87.4$&$128.7$&$28.0$\\
&$3^{+}$
&$18006^{\dagger}$
&$\sim0$&
&$\times$&&&&&
&$\times$\\
\midrule[1pt]
&&&\multicolumn{2}{c}{$bbb{\otimes}ccs$}&\multicolumn{6}{c}{$bbc{\otimes}cbs$}&\multicolumn{4}{c}{$bbs{\otimes}ccb$}\\
\cmidrule(lr){4-5}
\cmidrule(lr){6-11}
\cmidrule(lr){12-15}
System&$J^{P}$&Mass
&$\Omega_{bbb}^{*}\Omega_{cc}^{*}$
&$\Omega_{bbb}^{*}\Omega_{cc}$
&$\Omega_{bbc}^{*}\Omega_{cb}^{*}$
&$\Omega_{bbc}^{*}\Omega_{cb}$
&$\Omega_{bbc}^{*}\Omega_{cb}'$
&$\Omega_{bbc}\Omega_{cb}^{*}$
&$\Omega_{bbc}\Omega_{cb}$
&$\Omega_{bbc}\Omega_{cb}'$
&$\Omega_{bb}^{*}\Omega_{ccb}^{*}$
&$\Omega_{bb}^{*}\Omega_{ccb}$
&$\Omega_{bb}\Omega_{ccb}^{*}$
&$\Omega_{bb}\Omega_{ccb}$\\
\midrule[1pt]
$bbbccs$&$0^{+}$
&$18109^{\dagger}$
&$\times$&
&$\times$&&&&$\times$&$\times$
&$\times$&&&$\times$\\
&
&$18251$
&$8.9$&
&$\times$&&&&$129.7$&$0.1$
&$\times$&&&$2.6$\\
&
&$18362$
&$8.5$&
&$130.1$&&&&$10.6$&$17.0$
&$251.8$&&&$28.4$\\
&$1^{+}$
&$18040^{\dagger}$
&$\times$&$\times$
&$\times$&$\times$&$\times$&$\times$&$\times$&$\times$
&$\times$&$\times$&$\times$&$\times$\\
&
&$18110^{\dagger}$
&$\times$&$0.2$
&$\times$&$\times$&$\times$&$\times$&$\times$&$\times$
&$\times$&$\times$&$\times$&$\times$\\
&
&$18256$
&$4.4$&$6.3$
&$\times$&$64.4$&$8.5$&$21.9$&$15.7$&$25.7$
&$\times$&$\times$&$\times$&$9.8$\\
&
&$18323$
&$7.0$&$0.02$
&$75.2$&$14.2$&$15.5$&$0.2$&$0.04$&$18.7$
&$29.3$&$9.9$&$96.8$&$105.4$\\
&
&$18328$
&$0.1$&$1.5$
&$10.7$&$26.0$&$5.5$&$48.0$&$0.9$&$14.5$
&$165.0$&$39.8$&$71.5$&$0.1$\\
&$2^{+}$
&$18041^{\dagger}$
&$\times$&$\times$
&$\times$&$\times$&$\times$&$\times$&&
&$\times$&$\times$&$\times$\\
&
&$18112^{\dagger}$
&$\times$&$0.01$
&$\times$&$\times$&$\times$&$\times$&&
&$\times$&$\times$&$\times$\\
&
&$18308$
&$1.7$&$1.2$
&$30.6$&$2.4$&$37.2$&$8.0$&&
&$84.0$&$124.7$&$24.6$\\
&$3^{+}$
&$18112^{\dagger}$
&$\times$&
&$\times$&&&&&
&$\times$\\
\bottomrule[1pt]
\bottomrule[1pt]
\end{tabular}
\end{table*}
%
\begin{table*}[htbp]
\centering
\caption{The partial width ratios for the $bbbccq$ dibaryon states. For each state, we choose one mode as the reference channel, and the partial width ratios of the other channels are calculated relative to this channel. For a dibaryon state decays into two channels through two different quark rearrangements, the $\gamma_{i}$'s, which depend on the wave functions of final states, may not equal (or approximately equal). Thus we compare only the decay width ratios between channels through same quark rearrangement type. The scattering states are marked with a dagger ($\dagger$). The masses are all in unit of MeV.}
\label{table:bbbccq:decay:Ratio}
\begin{tabular}{cccccccccccccccccccccc}
\toprule[1pt]
\toprule[1pt]
&&&\multicolumn{2}{c}{$bbb{\otimes}ccn$}&\multicolumn{6}{c}{$bbc{\otimes}cbn$}&\multicolumn{4}{c}{$bbn{\otimes}ccb$}\\
\cmidrule(lr){4-5}
\cmidrule(lr){6-11}
\cmidrule(lr){12-15}
System&$J^{P}$&Mass
&$\Omega_{bbb}^{*}\Xi_{cc}^{*}$
&$\Omega_{bbb}^{*}\Xi_{cc}$
&$\Omega_{bbc}^{*}\Xi_{cb}^{*}$
&$\Omega_{bbc}^{*}\Xi_{cb}$
&$\Omega_{bbc}^{*}\Xi_{cb}'$
&$\Omega_{bbc}\Xi_{cb}^{*}$
&$\Omega_{bbc}\Xi_{cb}$
&$\Omega_{bbc}\Xi_{cb}'$
&$\Xi_{bb}^{*}\Omega_{ccb}^{*}$
&$\Xi_{bb}^{*}\Omega_{ccb}$
&$\Xi_{bb}\Omega_{ccb}^{*}$
&$\Xi_{bb}\Omega_{ccb}$\\
\midrule[1pt]
$bbbccn$&$0^{+}$
&$18003^{\dagger}$
&$\times$&
&$\times$&&&&$\times$&$\times$
&$\times$&&&$\times$\\
&
&$18167$
&$1.0$&
&$\times$&&&&$79.6$&$1$
&$\times$&&&$1$\\
&
&$18280$
&$1.0$&
&$8.4$&&&&$1.3$&$1$
&$8.5$&&&$1$\\
&$1^{+}$
&$17941^{\dagger}$
&$\times$&$\times$
&$\times$&$\times$&$\times$&$\times$&$\times$&$\times$
&$\times$&$\times$&$\times$&$\times$\\
&
&$18004^{\dagger}$
&$\times$&$1$
&$\times$&$\times$&$\times$&$\times$&$\times$&$\times$
&$\times$&$\times$&$\times$&$\times$\\
&
&$18161$
&$1$&$1.3$
&$\times$&$6.5$&$1$&$2.4$&$1.4$&$3.6$
&$\times$&$\times$&$\times$&$1$\\
&
&$18233$
&$1$&$\sim0$
&$31.0$&$4.1$&$7.3$&$1$&$0.1$&$3.5$
&$1$&$0.4$&$22.4$&$17.4$\\
&
&$18250$
&$1$&$3.0$
&$1.6$&$26.2$&$1$&$32.8$&$1.3$&$13.1$
&$27.8$&$6.9$&$5.7$&$1$\\
&$2^{+}$
&$17942^{\dagger}$
&$\times$&$\times$
&$\times$&$\times$&$\times$&$\times$&&
&$\times$&$\times$&$\times$\\
&
&$18006^{\dagger}$
&$\times$&$1$
&$\times$&$\times$&$\times$&$\times$&&
&$\times$&$\times$&$\times$\\
&
&$18230$
&$0.2$&$1$
&$22.1$&$1$&$25.2$&$5.7$&&
&$3.1$&$4.6$&$1$\\
&$3^{+}$
&$18006^{\dagger}$
&$1$&
&$\times$&&&&&
&$\times$\\
\midrule[1pt]
&&&\multicolumn{2}{c}{$bbb{\otimes}ccs$}&\multicolumn{6}{c}{$bbc{\otimes}cbs$}&\multicolumn{4}{c}{$bbs{\otimes}ccb$}\\
\cmidrule(lr){4-5}
\cmidrule(lr){6-11}
\cmidrule(lr){12-15}
System&$J^{P}$&Mass
&$\Omega_{bbb}^{*}\Omega_{cc}^{*}$
&$\Omega_{bbb}^{*}\Omega_{cc}$
&$\Omega_{bbc}^{*}\Omega_{cb}^{*}$
&$\Omega_{bbc}^{*}\Omega_{cb}$
&$\Omega_{bbc}^{*}\Omega_{cb}'$
&$\Omega_{bbc}\Omega_{cb}^{*}$
&$\Omega_{bbc}\Omega_{cb}$
&$\Omega_{bbc}\Omega_{cb}'$
&$\Omega_{bb}^{*}\Omega_{ccb}^{*}$
&$\Omega_{bb}^{*}\Omega_{ccb}$
&$\Omega_{bb}\Omega_{ccb}^{*}$
&$\Omega_{bb}\Omega_{ccb}$\\
\midrule[1pt]
$bbbccs$&$0^{+}$
&$18109^{\dagger}$
&$\times$&
&$\times$&&&&$\times$&$\times$
&$\times$&&&$\times$\\
&
&$18251$
&$1$&
&$\times$&&&&$1$&$\sim0$
&$\times$&&&$1$\\
&
&$18362$
&$1$&
&$12.2$&&&&$1$&$1.6$
&$8.9$&&&$1$\\
&$1^{+}$
&$18040^{\dagger}$
&$\times$&$\times$
&$\times$&$\times$&$\times$&$\times$&$\times$&$\times$
&$\times$&$\times$&$\times$&$\times$\\
&
&$18110^{\dagger}$
&$\times$&$1$
&$\times$&$\times$&$\times$&$\times$&$\times$&$\times$
&$\times$&$\times$&$\times$&$\times$\\
&
&$18256$
&$1$&$1.4$
&$\times$&$7.6$&$1$&$2.6$&$1.9$&$3.0$
&$\times$&$\times$&$\times$&$1$\\
&
&$18323$
&$1$&$\sim0$
&$4.9$&$0.9$&$1$&$0.01$&$\sim0$&$1.2$
&$3.0$&$1$&$9.8$&$10.7$\\
&
&$18328$
&$0.1$&$1$
&$1.9$&$4.7$&$1$&$8.7$&$0.2$&$2.6$
&$4.1$&$1$&$1.8$&$\sim0$\\
&$2^{+}$
&$18041^{\dagger}$
&$\times$&$\times$
&$\times$&$\times$&$\times$&$\times$&&
&$\times$&$\times$&$\times$\\
&
&$18112^{\dagger}$
&$\times$&$1$
&$\times$&$\times$&$\times$&$\times$&&
&$\times$&$\times$&$\times$\\
&
&$18308$
&$1.4$&$1$
&$12.6$&$1$&$15.2$&$3.3$&&
&$3.4$&$5.1$&$1$\\
&$3^{+}$
&$18112^{\dagger}$
&$\times$&
&$\times$&&&&&
&$\times$\\
\bottomrule[1pt]
\bottomrule[1pt]
\end{tabular}
\end{table*}
%

\bibliography{myreference}
\end{document}